\documentclass[%
 reprint,
 amsmath,amssymb,
 aps,
prb,
]{revtex4-1}

\usepackage{graphicx}
\usepackage{dcolumn}
\usepackage{bm}
\usepackage{epstopdf}

\begin{document}

\preprint{APS/123-QED}

\title{Multiband Effect and Possible Dirac Fermions in Fe$_{1+y}$Te$_{0.6}$Se$_{0.4}$}

\author{Yue Sun,$^{1,2}$ Toshihiro Taen,$^2$ Tatsuhiro Yamada,$^2$ Sunseng Pyon,$^2$ Terukazu Nishizaki,$^3$ Zhixiang Shi$^1$}
\email[]{zxshi@seu.edu.cn}
\author{Tsuyoshi Tamegai$^2$}
\email[]{ tamegai@ap.t.u-tokyo.ac.jp}

\affiliation{%
 $^{1}$Department of Physics and Key Laboratory of MEMS of the Ministry of Education, Southeast University, Nanjing 211189, People's Republic of China \\
$^2$Department of Applied Physics, The University of Tokyo, 7-3-1 Hongo, Bunkyo-ku, Tokyo 113-8656, Japan}
\affiliation{
$^{3}$Institute for Materials Research, Tohoku University, Sendai 980-8577, Japan
}
\altaffiliation{present address: Department of Electrical Engineering and Information Technology, Kyushu Sangyo University, Fukuoka 813-8503, Japan}
\date{\today}

\begin{abstract}
We investigated the transport properties of Fe$_{1+y}$Te$_{0.6}$Se$_{0.4}$ single crystals with different amounts of excess Fe prepared by O$_2$ annealing. The O$_2$ annealing remarkably improves transport properties. In particular, a strongly nonlinear Hall resistivity was observed only in the fully-annealed crystal, and the magnetoresistance (MR) is drastically enhanced after annealing, reaching a value larger than 17\% at 16 K and 14 T. The obvious change of transport properties after the annealing indicates that the band structure of Fe$_{1+y}$Te$_{0.6}$Se$_{0.4}$ is affected by the excess Fe. The nonlinear Hall resistivity and violation of (modified) Kohler's scaling of the large MR prove the multiband effects in the Fe$_{1+y}$Te$_{0.6}$Se$_{0.4}$ single crystal. The MR for the fully-annealed crystal develops linearly against magnetic field from intermediate field (e. g. 2 T at 16 K) to the measurement limit of 14 T. This phenomenon is interpreted by the existence of Dirac cone state, in which all the Dirac fermions occupy only the lowest Landau level in the quantum limit.

\begin{description}
\item[PACS numbers]
\verb+74.70.Xa+, \verb+74.25.F-+, \verb+72.15.Gd+, \verb+75.47.-m+

\end{description}
\end{abstract}

\pacs{Valid PACS appear here}
\maketitle
\section{introduction}
Recently discovered iron-based superconductors (IBSs) with superconducting transition temperature \emph{T}$_c$ above 55 K is another member of the high temperature superconductors (HTSs) after cuprate superconductors.\cite{KamiharaJACS,StewartIBSsreview} Although IBSs share some similarities with cuprate superconductors like layered structure, very high upper critical fields, and doping phase diagram, important differences exist between the two families. Cuprates are doped Mott insulators with strong correlation and a single band behavior, while IBSs are metallic with multiband electronic structures.\cite{StewartIBSsreview} Existence of disconnected Fermi surfaces with  electron and hole characters, and spin or orbital fluctuations are supposed to be responsible for the high value of \emph{T}$_c$  in IBSs based on either  \emph{s}$_\pm$\cite{MazinS} or \emph{s}$_{++}$ scenario\cite{KontaniPRL}. The nesting between electron and hole bands is supposed to be related to the high value of \emph{T}$_c$ in IBSs based on the scenario of \emph{s}$^\pm$ paring. This multiband feature also influences the normal state transport properties of IBSs. Strong temperature dependent Hall coefficients, large magnetoresistance (MR), and linear temperature dependence of resistivity were observed, especially in iron pnicitide.\cite{RullierPRL,OhgushiPRB,ChengPRBNd1111,SunJAP} Moreover, a Dirac cone state, coming from the nodes of the spin-density-wave (SDW) gap by complex zone folding in different bands, is observed in Ba/SrFe$_2$As$_2$ and La/PrFeAsO.\cite{RichardPRL,HuynhPRL,PallecchiPRB,ChongEPL,BhoiPrFeAsO} Although the weight of Dirac cone state is small, it can dominate the transport properties because of its extremely high mobility. In consequence, a large and linear temperature dependent MR was observed.

In the family of IBSs, iron chalcogenide Fe$_{1+y}$Te$_{1-x}$Se$_{x}$ attracted much attention due to its simple structure, which is convenient to probe the superconducting mechanism. And its less toxic nature is also advantageous to application in iron-based superconductors. Band structure calculations and angle-resolved photoemission spectroscopy (ARPES) prove the multiband structure in Fe$_{1+y}$(Te/Se),\cite{SubediPRB,XiaPRL,ChenPRB11} which is characterized by hole bands around $\Gamma$ point and electron bands around M point, similar to iron pnictides. However, the transport features characteristic of multiband structure like the large magnetoresistance has not been reported yet. Instead, a very small value and sometimes even negative MR was observed in previous reports.\cite{ChangSUST,HuPRB} This unexpected transport property may come from the Fe nonstoichiometries,\cite{BaoWeiPRL,BendelePRB} which originate from the partial occupation of excess Fe at the interstitial site in the Te/Se layer. The excess Fe is strongly magnetic, which provides local moments that interact with the adjacent Fe layers.\cite{ZhangPRB} In the parent compound Fe$_{1+y}$Te, the long-rang ($\pi$, 0) order can be tuned from commensurate to incommensurate by changing the amount of excess Fe.\cite{BaoWeiPRL} In Se doped superconducting Fe$_{1+y}$Te$_{1-x}$Se$_{x}$, neutron scattering measurements revealed that the excess Fe induces a magnetic Friedel-like oscillation that diffracts at ($\pi$, 0) order and involves more than 50 neighboring Fe sites.\cite{ThampyPRL} And the magnetic moment from excess Fe will also localize the charge carriers affecting the normal state transport properties.\cite{LiuPRB,SunAPRE} Thus, transport measurements on high-quality Fe$_{1+y}$Te$_{1-x}$Se$_{x}$ single crystal without the influence of excess Fe is crucial to reveal the intrinsic properties of iron chalcogenide, and is also helpful to the understanding of band structure and paring mechanism of IBSs.

In this paper, we benefit from the high-quality Fe$_{1+y}$Te$_{0.6}$Se$_{0.4}$ single crystals obtained by post-annealing to take accurate transport measurements.\cite{SunSUST,SunAPRE,Sunjpsj,SunJPSJshort,SunEPL,SunSciRep} The annealing remarkably improves transport properties. In particular, a nonlinear Hall resistivity was observed only in the fully-annealed crystal, and the magnetoresistance is drastically enhanced after annealing, reaching a value larger than 17\% at 16 K and 14 T, which proves the multiband property of Fe$_{1+y}$Te$_{0.6}$Se$_{0.4}$. Besides, a temperature dependent linear MR is observed in the annealed crystal, which is interpreted by the possible existence of Dirac fermions.
\section{experiment}
Single crystals with a nominal composition FeTe$_{0.6}$Se$_{0.4}$ was grown by the self-flux method.\cite{TaenPRB, SunSUST} The as-grown crystals were further annealed with controlled  amount of O$_2$ to partially (molar ratio \emph{n}(O):\emph{n}(sample)=0.7\%) or totally (\emph{n}(O):\emph{n}(sample)=1.5\%) remove the excess Fe to obtain the half-annealed or fully-annealed crystals, respectively. Combined Inductively-coupled plasma (ICP) atomic emission spectroscopy and Scanning tunneling microscopy (STM) measurements prove that the amount of excess Fe, \emph{y}, in the as-grown, half-annealed and fully-annealed crystals are roughly 0.14, 0.065 and 0, respectively. Details of the sample preparing and the composition analysis have been reported in our previous publications.\cite{SunSciRep} Magnetization measurements were performed using a commercial SQUID magnetometer (MPMS-XL5, Quantum Design). The Hall resistivity $\rho$$_{yx}$ and magnetoresistance $\rho$$_{xx}$ were measured at the same time using the six-lead method with the applied field parallel to \emph{c}-axis and perpendicular to the applied current. In order to decrease the contact resistance, we sputtered gold on the contact pads just after the cleavage, then attached gold wires on the contacts with silver paste. The Hall (MR) resistivity $\rho$$_{yx}$ ($\rho$$_{xx}$) was extracted from the difference (sum) of transverse (longitudinal) resistance measured at positive and negative fields, i.e., $\rho$$_{yx}$(\emph{H})=[$\rho$$_{yx}$(+\emph{H})-$\rho$$_{yx}$(-\emph{H})]/2 and $\rho$$_{xx}$(\emph{H})=[$\rho$$_{xx}$(+\emph{H})+$\rho$$_{xx}$(-\emph{H})]/2, which can effectively eliminate the longitudinal (transverse) resistivity component due to the misalignment of contacts.
\begin{figure}\center

　　\includegraphics[width=8.5cm]{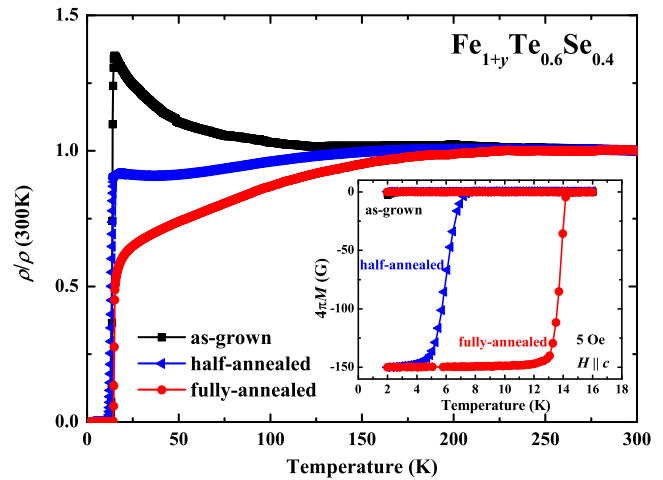}\\
　　\caption{(Color online) Temperature dependence of the resistivities scaled by the values at 300 K for the as-grown, half-annealed and fully-annealed Fe$_{1+y}$Te$_{0.6}$Se$_{0.4}$ single crystals. The inset shows the temperature dependences of ZFC and FC magnetizations at 5 Oe for the three samples.}\label{}
\end{figure}
\section{results and discussion}
Inset of Fig. 1 shows the temperature dependence of zero-field-cooled (ZFC) and field-cooled (FC) magnetization at 5 Oe for the as-grown, half-annealed and fully-annealed Fe$_{1+y}$Te$_{0.6}$Se$_{0.4}$ single crystals. The as-grown crystal usually shows no superconductivity or very weak diamagnetic signal below 3 K. The very low \emph{T}$_c$ is usually attributed to the existence of excess Fe in the interstitial sites. Density functional study shows that the excess Fe in the interstitial site is magnetic and interacts with magnetism of Fe in the Fe planes.\cite{ZhangPRB} The magnetic moment from excess Fe will act as a pair breaker, so that the superconductivity is almost totally suppressed in the as-grown crystal. After partially removing the excess Fe by O$_2$ annealing, the half-annealed crystal shows \emph{T}$_c$ $\sim$ 7.5 K with transition width about 1.5 K (obtained from the criteria of 10 and 90\% of the magnetization). This relatively sharp transition width indicates that the left excess Fe is almost homogeneously distributed in the sample. After totally removing the excess Fe by O$_2$ annealing, the fully-annealed crystal shows \emph{T}$_c$ $\sim$ 14.3 K with the transition width less than 1 K. In the main panel of Fig. 1, we compared the temperature dependence of resistivities, scaled by the values at 300 K. Resistivities for all the three crystals maintain a nearly constant value above 150 K. From 150 K down to the superconducting transition temperature, the as-grown sample shows a nonmetallic behavior (d$\rho$/d\emph{T} $<$ 0). This nonmetallic behavior was suppressed by removing the excess Fe and a flattened resistive behavior above \emph{T}$_c$ was found in the half-annealed crystal. When the excess Fe is totally removed as in the fully-annealed crystal, resistivity manifests a metallic behavior (d$\rho$/d\emph{T} $>$ 0). The divergence in resistivity below 150 K for the three crystals is also caused by the magnetic moment from excess Fe, which will localize the charge carriers and increase the resistivity.\cite{LiuPRB,SunAPRE} Such localization effect from excess Fe will be studied in  detail later. The resistive results for the as-grown and half-annealed crystals show higher \emph{T}$_c$ compared with that from magnetization measurements, which is coming from the filamentary superconductivity. Actually, no specific heat jump associated with superconducting transition can be observed in the as-grown crystal.\cite{SunSciRep} The filamentary superconductivity may come from some small parts of the crystal containing less amount of excess Fe, like the surface layers, which may have been slightly annealed at room temperature in the air.\cite{DongPRB}
\begin{figure}\center

　　\includegraphics[width=8.5cm]{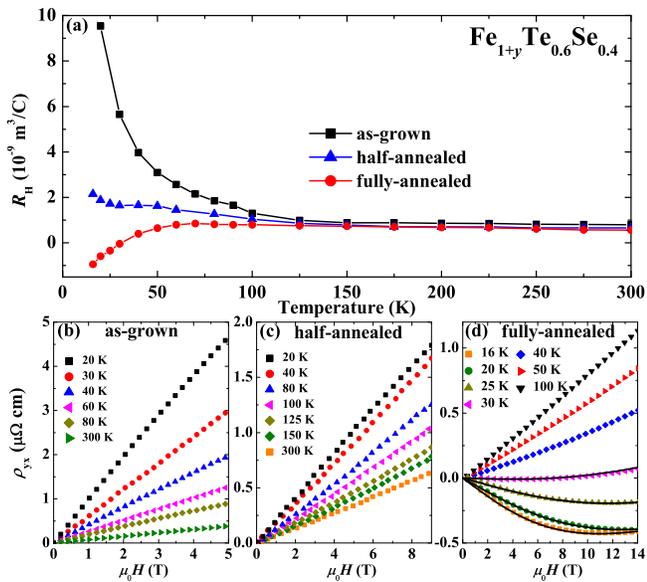}\\
　　\caption{(Color online) (a) Hall coefficients \emph{R}$_H$ for the as-grown, half-annealed and fully-annealed Fe$_{1+y}$Te$_{0.6}$Se$_{0.4}$ single crystals. Hall resistivity $\rho$$_{xy}$ at several temperatures for the (b) as-grown, (c) half-annealed, and (d) fully-annealed crystals.}\label{}
\end{figure}

Figs. 2(b)-(d) show the Hall resistivity $\rho$$_{yx}$ at several temperatures for the as-grown, half-annealed, and fully-annealed Fe$_{1+y}$Te$_{0.6}$Se$_{0.4}$ single crystals, respectively. The $\rho$$_{yx}$ for the as-grown crystal follows a linear relationship with the applied field and has a positive slope, d$\rho$$_{yx}$/d\emph{H} $>$ 0. Also, the values of $\rho$$_{yx}$ at different temperatures above \emph{T}$_c$ are all positive, indicating the electrical transport is dominated by hole-type carriers. For the half-annealed crystal, $\rho$$_{yx}$ still keeps positive and  linearly increases with magnetic field. However, $\rho$$_{yx}$ of the fully-annealed crystal becomes negative when temperature decreases below 40 K, and an obvious nonlinear behavior can be witnessed. The nonlinear behavior and sign reversal observed in $\rho$$_{yx}$ proves the existence of multiband effect. Similar behavior of the $\rho$$_{yx}$ has been also observed in FeSe single crystal and FeTe$_{0.5}$Se$_{0.5}$ thin films.\cite{LeiPRB,TsukadaPRB} Hall coefficients \emph{R}$_H$ can be simply obtained from \emph{R}$_H$ = $\rho$$_{yx}$/\emph{$\mu$}$_0$\emph{H}, and were shown in Fig. 2(a). For the nonlinear $\rho$$_{yx}$ at low temperatures in the fully-annealed crystal, \emph{R}$_H$ was simply calculated from the linearly part at small fields. \emph{R}$_H$ is almost temperature independent above 100 K, and keeps a constant value $\sim$1 $\times$ 10$^{-9}$ m$^3$/C for all the three samples. When temperature decreases below 100 K, an obvious divergence in \emph{R}$_H$ is observed. In the as-grown crystal, \emph{R}$_H$  gradually increases with decreasing temperature showing an obvious upturn at low temperatures. This upturn is almost suppressed in the half-annealed crystal, in which \emph{R}$_H$ just slightly increases with decreasing temperature. In the fully-annealed crystal, \emph{R}$_H$ keeps nearly temperature independence above 60 K, followed by a sudden decrease, and finally changes sign from positive to negative before approaching \emph{T}$_c$. The sign reversal in Hall coefficient is usually attributed to the multiband structure, indicating the dominance of electron in the charge carriers before the occurrence of superconductivity in Fe$_{1+y}$Te$_{0.6}$Se$_{0.4}$.
\begin{figure}\center

　　\includegraphics[width=8.5cm]{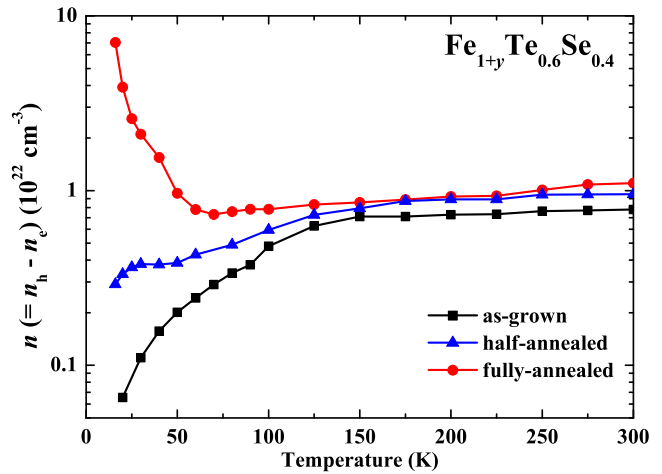}\\
　　\caption{(Color online) Temperature dependence of the ``charge carrier densities'' \emph{n}(=\emph{n}$_h$-\emph{n}$_e$) for the as-grown, half-annealed and fully-annealed crystals.}\label{}
\end{figure}

Actually, multiband structure has already been proven to be a common property in IBSs. According to the band calculations and  angle-resolved photoemission spectroscopy (ARPES) results, at least four bands originated from Fe 3\emph{d} orbitals cross the Fermi level.\cite{SubediPRB,XiaPRL} Two of them contribute hole-type charge carriers, and the other two contribute electron-type charge carriers. To quantitatively study the multiband effect in Fe$_{1+y}$Te$_{0.6}$Se$_{0.4}$, we first apply a simplified two-carrier model including one electron band with electron density \emph{n}$_e$ and mobility \emph{$\mu$}$_e$, and one hole type band with hole density \emph{n}$_h$ and mobility \emph{$\mu$}$_h$. According to the classical expression for the Hall coefficient of two-band model,\cite{SmithBook}
\begin{equation}
\label{eq.1}
R_H=\frac{1}{e}\frac{(\mu_h^2n_h-\mu_e^2n_e)+(\mu_h\mu_e)^2(\mu_0H)^2(n_h-n_e)}{(\mu_en_e+\mu_hn_h)^2+(\mu_h\mu_e)^2(\mu_0H)^2(n_h-n_e)^2}.
\end{equation}
The field dependence of $\rho$$_{yx}$ will become nonlinear when the densities of electrons and holes are different. For the as-grown and half-annealed crystals, $\rho$$_{yx}$ is positive and almost linear in field at temperature from 20 to 300 K, indicating the hole-type carrier is dominant. The hole density can be simply obtained by \emph{n}$_h$ = 1/e\emph{R}$_H$. In the fully-annealed crystal, hole-type carrier is still dominant at high temperatures. However, $\rho$$_{yx}$ exhibits obvious nonlinear behavior below 40 K and even changes sign to negative at temperature below 30 K. The nonlinear behavior is a signature of the coexistence of electron- and hole-type carriers, and can be well fitted by Eq. (1) as shown with the solid line in Fig. 2(d). The obtained ``charge carrier densities'' \emph{n} = \emph{n}$_h$ - \emph{n}$_e$ in the as-grown, half-annealed and fully-annealed crystals are shown in Fig. 3. It is obvious that ``charge carrier densities'' for all the three samples keep almost temperature independence above 100 K. Below this temperature, \emph{n} of the as-grown crystal reduces quickly with decreasing temperature. When the excess Fe was partially removed by annealing, the reduction of \emph{n} is obviously suppressed in the half-annealed crystal. Furthermore, in the fully-annealed crystal without excess Fe, \emph{n} increases quickly with decreasing temperature below 70 K. Moreover, Eq. (1) also predicts that \emph{R}$_H$ = \emph{e}$^{-1}$(\emph{n}$_h$$\mu$${_h^2}$-\emph{n}$_e$$\mu$$_e^2$)/(\emph{n}$_e$$\mu$$_e$+\emph{n}$_h$$\mu$$_h$)$^2$ when $\mu$$_0$\emph{H} $\rightarrow$ 0, and \emph{R}$_H$ = \emph{e}$^{-1}$ $\times$ 1/(\emph{n}$_h$-\emph{n}$_e$) when $\mu$$_0$\emph{H} $\rightarrow$ $\infty$.  For the fully-annealed crystal, the slope of $\rho$$_{yx}$ at temperatures below 30 K changes sign from negative at low fields to positive at high fields, which means (\emph{n}$_h$$\mu$${_h^2}$-\emph{n}$_e$$\mu$$_e^2$) $<$ 0 and \emph{n}$_h$-\emph{n}$_e$ $>$ 0. This means that $\mu$$_e$ $>$ $\mu$$_h$ at low temperatures. It shows that although the hole density increases at low temperatures, the multiband effect becomes dominant because of the contribution of electron band with higher mobility. Until now, the classical two-band model successfully explains the nonlinear behavior of $\rho$$_{yx}$ and the multiband effect. However, the large increase of ``charge carrier density'' \emph{n} with decreasing temperature seems very unphysics because no obvious band structure change or opening of energy gap were reported. These results may just come from the emergence of a small band with very high mobility like that reported Dirac cone state in BaFe$_2$As$_2$, which will be discussed later. The evident differences in charge carrier densities of the three crystals indicate that the band structure of Fe$_{1+y}$Te$_{0.6}$Se$_{0.4}$ may change after annealing. Before annealing, the hole bands dominate the electronic transport of the as-grown crystal, and the contribution from electron bands is almost negligible. After annealing, the magnitude of the hole bands changes little since the Hall coefficient at higher temperature is close to that of the as-grown one. However, the electron bands emerge and contribute notably to the electronic transport at temperature below 150 K in the fully-annealed crystal. Actually, the band structure change after annealing was also witnessed by ARPES on single-layer FeSe film.\cite{HeNatMat}
\begin{figure}\center

　　\includegraphics[width=8.5cm]{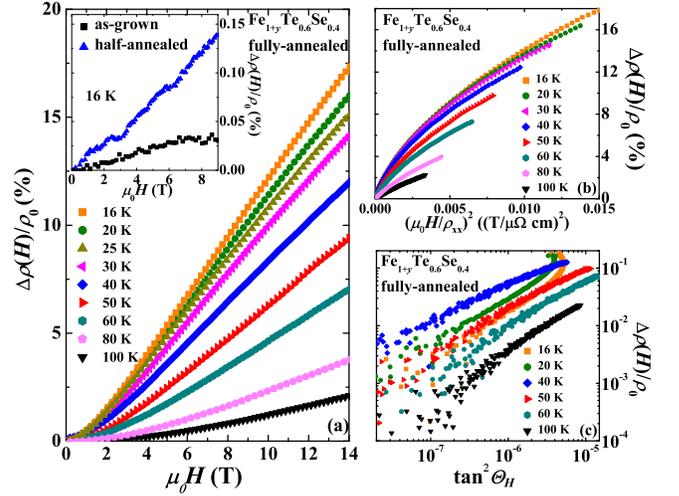}\\
　　\caption{(Color online) (a) Magnetic field dependence of  magnetoresistance (MR=($(\rho(H)-\rho(0))/\rho(0)$) for the fully-annealed Fe$_{1+y}$Te$_{0.6}$Se$_{0.4}$ single crystal at different temperatures. Inset is the MR for the as-grown and half-annealed crystals at 16 K. (b) MR for the fully-annealed crystal plotted as a function of $(\mu_0H/\rho_{xx})^2$. (c) MR for the fully-annealed crystal plotted as a function of tan$^2$$\Theta_H$.}\label{}
\end{figure}

To further investigate the multiband effect of Fe$_{1+y}$Te$_{0.6}$Se$_{0.4}$, we also studied the magnetoresistance of the three crystals. In a multiband system, the MR is usually described by the following expression,\cite{EomPRB}
\begin{equation}
\label{eq.2}
MR\equiv\frac{\Delta\rho(H)}{\rho(0)}\approx\frac{1}{2}\frac{\Sigma_i\Sigma_{j\neq i}\sigma_i\sigma_j(\omega_{ci}\tau_i-\omega_{cj}\tau_j)^2}{(\Sigma_i\sigma_i)^2},
\end{equation}
where $\sigma_i$ is the conductivity, $\tau_i$ is the relaxation time and $\omega_{ci}$ is the cyclotron frequency, which has opposite sign for electron and hole bands. In this case, the ($\omega_{ci}$$\tau_i$-$\omega_{cj}$$\tau_j$)$^2$ term becomes larger because the $\omega_{ci}$$\tau_i$ terms add up, which will results in a large MR. However, the MR of the as-grown crystal is just $\sim$ 0.03\% at 16 K under 9 T as shown in the inset of Fig. 4(a) . This unexpected small MR can be also explained by the effect of excess Fe. Recent neutron scattering measurements revealed that the excess Fe induces a magnetic Friedel-like oscillation at ($\pi$, 0) order and involves more than 50 neighboring Fe sites.\cite{ThampyPRL} Spins from those Fe clusters will be weakly polarized under magnetic field inducing a negative MR, which will cancel out the positive MR of the sample itself. Actually, previous reports on Fe$_{1+y}$Te$_{1-x}$Se$_x$ all show such small values of MR, and sometimes even negative MR was observed.\cite{ChangSUST,HuPRB} Such small MR is increased to $\sim$ 0.14\% in the half-annealed crystal because parts of the excess Fe was removed. After totally removing the excess Fe, MR of the fully-annealed Fe$_{1+y}$Te$_{0.6}$Se$_{0.4}$ reaches larger than 17\% at 16 K under 14 T. Such a large MR is observed in iron chalcogenide superconductors for the first time, which supports the multiband structure proved by ARPES and first principle calculation.\cite{SubediPRB,XiaPRL,ChenPRB11}  In the following, we will focus on the transport properties of the fully-annealed crystal, which has little influence from excess Fe and manifests the intrinsic property of FeTe$_{0.6}$Se$_{0.4}$.

In the conventional Fermi liquid state of a single-band system with isotropic scattering, the MR can be simply scaled by the Kohler's law,\cite{ZimanBook} $\Delta\rho(H)/\rho(0)=F(\omega_c\tau)=F[(\mu_0H/\rho(0))^2]$, where \emph{F} is a function of the cyclotron frequency $\omega_c$ and scattering time $\tau$. The scaling of the MR of the fully-annealed crystal is plotted in Fig. 4(b), which is obviously violating the Kohler's rule. Until now, we cannot simply attribute the violation of the Kohler's scaling to the multiband effect since the violation is also found in some strongly correlated materials like high \emph{T}$_c$-cuprates and heavy fermion intermetalics.\cite{HarrisPRL,NakajimaJPSJ} In these compounds, MR can be scaled by the modified Kohler's rule,\cite{KontaniRepProg} $\Delta\rho(H)/\rho(0)$$\propto$tan$^2$$\Theta_H$. To examine this relation, we also plot the MR as a function of  tan$^2$$\Theta_H$ in Fig.4 (c). Obviously, our data also violates the modified Kohler's rule. Thus, we attribute the violation of (modified) Kohler's rule to the multiband effect in Fe$_{1+y}$Te$_{0.6}$Se$_{0.4}$.
\begin{figure}\center

　　\includegraphics[width=8.5cm]{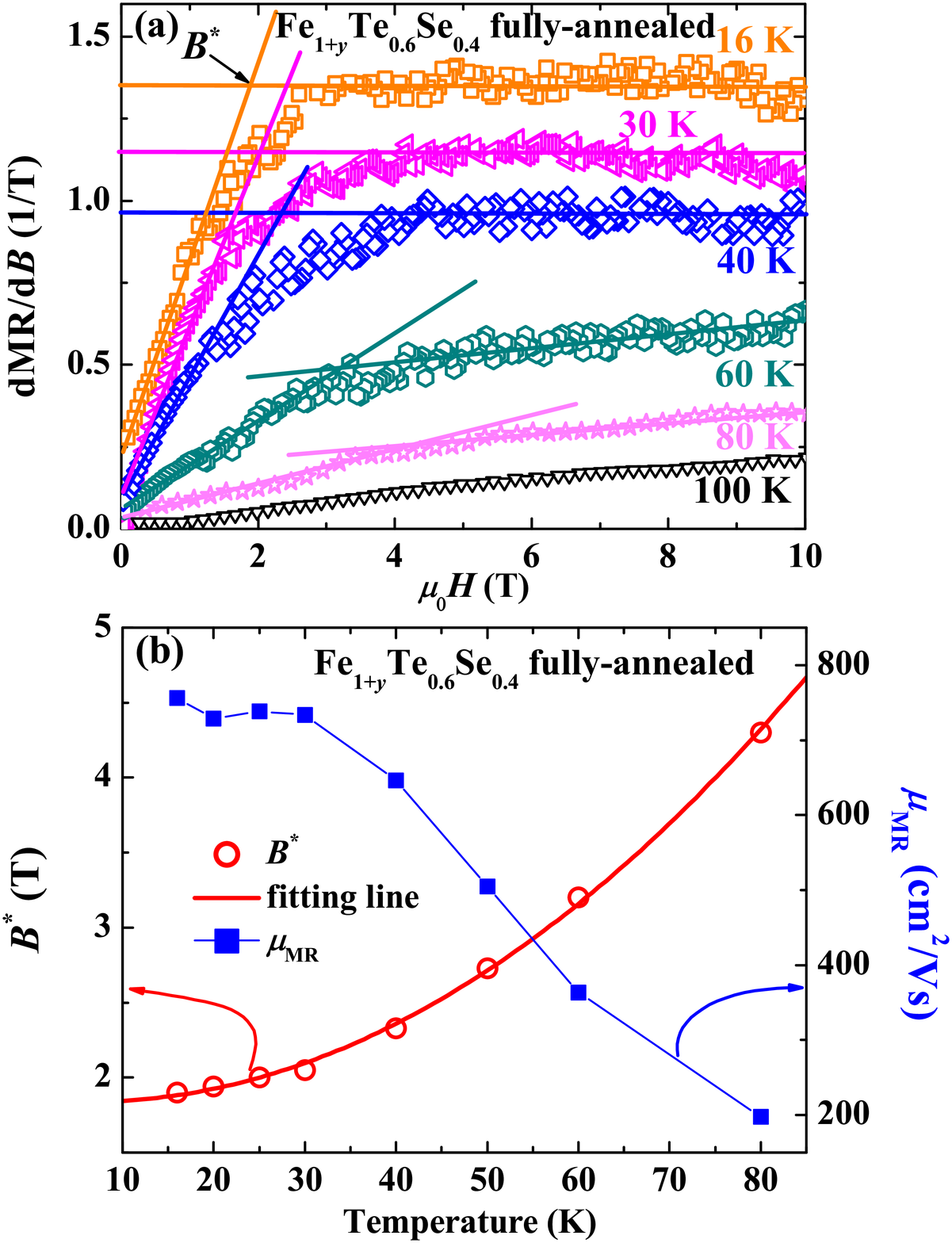}\\
　　\caption{(Color online) (a) The field derivative of in-plane MR at different temperatures for the fully-annealed Fe$_{1+y}$Te$_{0.6}$Se$_{0.4}$ single crystal. The solid lines denote the semiclassical regime and the quantum linear region, respectively. The characteristic field \emph{B}$^*$ is marked by the arrow. (b) Temperature dependence of the characteristic field \emph{B}$^*$ (red circles) and the effective MR mobility $\mu_{MR}$ (blue squares). Red line is the fitting of \emph{B}$^*$ by \emph{B}$^*$ = (1/2e$\hbar$\emph{v}$^2_F$)(k$_B$\emph{T}+\emph{E}$_F$)$^2$.}\label{}
\end{figure}

More interesting, MR of the fully-annealed Fe$_{1+y}$Te$_{0.6}$Se$_{0.4}$ linearly increases with the applied field from intermediate field (e. g. 2 T at 16 K) to the measurement limit of 14 T, whereas a small parabolic-like bend just remains at the low fields. This is in sharp contrast to the semiclassical quadratic field dependence of MR, in which MR generally develops in proportion to \emph{H}$^2$ over the entire field range. The linear dependence of MR on field is more evident in the first-order derivative d\emph{MR}/d\emph{B} as shown in Fig. 5(a). d\emph{MR}/d\emph{B} is proportional to magnetic field at low \emph{H}, then saturated at high fields. The linear MR can be interpreted by several possible mechanisms. For instance, in a sample with one-dimensional Fermi surface, although the MR shows quadratic field dependence along the open orbits, linear MR might be observed in the polycrystalline sample due to averaging effect.\cite{AbrikosovPRB} This mechanism is obviously not suitable for our single crystalline sample. The linear MR is also observed in heavily disordered system,\cite{ParishNature} which is not applied to Fe$_{1+y}$Te$_{0.6}$Se$_{0.4}$. The linear MR can be also interpreted by considering a quantum limit where all the carriers occupy only the lowest Laudau level (LL).\cite{AbrikosovPRB,AbrikosovEPL} This situation usually happens when the field is very large and the difference between the zeroth and first Landau levels $\Delta_{LL}$ exceeds the Fermi energy \emph{E$_F$} and the thermal fluctuation \emph{k}$_B$\emph{T}. In such a quantum limit, MR can no longer be described in the framework of the conventional Born scattering approximation like Eq. (2), and is instead expressed as:
\begin{equation}
\label{eq.3}
MR=\frac{1}{2\pi}(\frac{e^2}{\varepsilon_\infty\hbar v_F})^2\frac{N_i}{en^2}Bln(\varepsilon_\infty),
\end{equation}
where $N_i$ is the density of  scattering centers, \emph{n} is the carrier density, $v_F$ is the Fermi velocity and $\varepsilon_\infty$ is the high-frequency dielectric constant.\cite{AbrikosovPRB,AbrikosovEPL} In a conventional parabolic band, the LL is proportional to \emph{B}, $\Delta_{LL}$=e$\hbar$\emph{B}/\emph{m}$^*$, where \emph{m}$^*$ is the effective mass. To satisfy the quantum limit, \emph{i. e.}, $\Delta_{LL}>k_BT$,  a very large value of magnetic field is needed. Thus, the linear MR coming from the quantum limit is difficult to be observed in a moderate field range. By contrast, the linear MR was identified in low field region in some materials hosting Dirac fermions with linear energy dispersion, such as graphene,\cite{NovoselovNature} topological insulators,\cite{TaskinPRL} Ag$_{2-\delta}$(Te/Se),\cite{XuNature} $\alpha$-(BEDT-TTF)$_2$I$_3$,\cite{Kobayashijpsj} some layered compounds with two-dimensional Fermi surface (like SrMnBi$_2$)\cite{ParkPRL,WangPRB} and iron-based Ba(Sr)Fe$_2$As$_2$\cite{RichardPRL,HuynhPRL,ChongEPL}  and La(Pr)FeAsO.\cite{PallecchiPRB,BhoiPrFeAsO} For the Dirac state, $\Delta_{LL}$ is described as $\Delta_{LL}$=$\pm$\emph{v}$_F$$\sqrt{2e\hbar B}$, leading to a much larger LL splitting compared with the parabolic band. Consequently, the quantum limit  can be achieved in low field region.\cite{AbrikosovPRB}

Characteristic field \emph{B}$^*$, defined as the crossover field between the semiclassical regime and the quantum linear regime, is marked by the arrow in Fig. 5(a). The temperature dependence of \emph{B}$^*$ is shown in Fig. 5(b), which is obviously violating the linear relation expected from conventional parabolic bands, and can be well fitted by \emph{B}$^*$ = (1/2e$\hbar$\emph{v}$^2_F$)(k$_B$\emph{T}+\emph{E}$_F$)$^2$ for the Dirac fermions as shown in Fig.5 (b).\cite{TaskinPRL} The good agreement of \emph{B}$^*$ with the above equation confirms the existence of Dirac fermions in Fe$_{1+y}$Te$_{0.6}$Se$_{0.4}$. The fitting gives a large Fermi velocity \emph{v}$_F$ $\sim$ 1.1 $\times$ 10$^5$ ms$^{-1}$ and \emph{E}$_F$ $\sim$ 5.5 meV, which are close to the previous reports in similar compounds BaFe$_2$As$_2$ (\emph{v}$_F$ $\sim$ 1.88 $\times$ 10$^5$ ms$^{-1}$, \emph{E}$_F$ $\sim$ 2.48 meV)\cite{HuynhPRL} and SrFe$_2$As$_2$ (\emph{v}$_F$ $\sim$ 3.11 $\times$ 10$^5$ ms$^{-1}$ and \emph{E}$_F$ $\sim$ 6.9 meV).\cite{ChongEPL} If we focus on the data for fields close to zero, we can obtain the coefficient of the \emph{B}$^2$ quadratic term \emph{A}$_2$. In a multiband system with both Dirac and coventional parabolic band where Dirac carriers are dominant in transport, the coefficient \emph{A}$_2$ is related to the effective MR mobility $\sqrt{A_2}$ = $\frac{\sqrt{\sigma_e\sigma_h}}{\sigma_e+\sigma_h}$($\mu_e$+$\mu_h$) = $\mu_{MR}$.\cite{TaskinPRL,KuoPRB} The effective MR is smaller than the average mobility of carriers $\mu_{avg}$ = ($\mu_e+\mu_h$)/2, and gives an estimation of the lower bound. Temperature dependence of $\mu_{MR}$ was calculated and shown in Fig.5 (b).  The $\mu_{MR}$ reaches a value close to 800 cm$^2$/Vs at 16 K comparable to the value ($\sim$ 1000 cm$^2$/Vs) obtained from quantum oscillation in BaFe$_2$As$_2$.\cite{AnalytisPRB} The value of mobility obtained here is much larger than that reported in FeTe$_{1-x}$Se$_x$ thin films ($<$ 10 cm$^2$/Vs),\cite{TsukadaJPSJ} which again confirms the existence of Dirac fermions. $\mu_{MR}$ decreases with increasing temperature since thermal fluctuation smear out the LL splitting. The Dirac cone state is hardly observed when temperature increases above 80 K. If we look back to the part of Hall effect, the calculated value of \emph{n} is also drastically increased below 70 K, which is the same temperature region for the Dirac cone state becomes dominant. Thus, the steep increase of \emph{n} should be reinterpreted as a sign of the emergence of Dirac fermions.

The Dirac cone states have been theoretically predicted and experimentally confirmed by ARPES in BaFe$_2$As$_2$.\cite{RanPRB,RichardPRL} The formation of Dirac cone state in BaFe$_2$As$_2$ is a consequence of the nodes of the SDW gap by complex zone folding in bands with different parties,\cite{RichardPRL} and it can coexist with superconductivity in Ru-doped BaFe$_2$As$_2$ until the vanish of SDW.\cite{TanabePRBCoexis}\cite{TanabePRBSuppre} Similar results have also been reported in Ru-doped LaFeAsO,\cite{PallecchiPRB} which seems to indicate that the emergency of Dirac fermions in iron pnictides is usually accompanying with SDW. However, in iron-chalcogenide FeTe$_{0.6}$Se$_{0.4}$, the long-range SDW is already totally suppressed according to the phase diagram,\cite{LiuNatMat,KawasakiSSC} which indicates the origin of the Dirac cone state in iron-chalcogenides may be different from that in iron pnictides. Here, we should point out that although the long-range SDW is proven to be suppressed by Se doping, the SDW fluctuation may still survive. So we cannot simply eliminate the possible origin of the Dirac cone state coming from the behavior of SDW fluctuation. On the other hand, band structure calculation on FeTe/Se shows a linear bands crossover near the Fermi surface around M point.\cite{MiyakeJPSJ} It also manifests that the Dirac cone like structure is in electronic state, which is consistent with our transport results. The missing of Dirac cone states in reported ARPES results may be attributed to the effects of remaining excess Fe, which will change the band structure and also localize the charge carriers. Thus, future experiments like ARPES or Shubnikov-de Hass oscillations on the fully-annealed FeTe$_{0.6}$Se$_{0.4}$ crystal without excess Fe is promising to reveal the origin of the Dirac cone state.
\section{conclusions}
In summary, we performed detailed investigations of transport properties on Fe$_{1+y}$Te$_{0.6}$Se$_{0.4}$ single crystals with different amounts of excess Fe. The semiconducting resistive behavior in the as-grown crystal was gradually suppressed by annealing in O$_2$, and was replaced by a metallic behavior in the fully-annealed crystal. The as-grown and half-annealed crystals show a linear Hall resistivity and very small value of MR. On the other hand, the fully-annealed crystal manifests an obvious nonlinear $\rho$$_{yx}$ and large MR, which is over 17\% at 16 K under 14 T. The nonlinear $\rho$$_{yx}$ and large MR together with the violation of (modified) Kohler's rule prove the multiband effect in FeTe$_{0.6}$Se$_{0.4}$. The comparison of the transport properties for the as-grown, half-annealed, and fully-annealed Fe$_{1+y}$Te$_{0.6}$Se$_{0.4}$ single crystals indicates that the band structure change after O$_2$ annealing. The MR for the fully-annealed crystal also shows a linear increase against magnetic field from intermediate field to the measurement limit, which is interpreted by the existence of Dirac fermions.

\acknowledgements
We thank Akira Sugimoto and Toshikazu Ekino for the STM measurements, and Ryotaro Arita for helpful discussion. This work was partly supported by the Natural Science Foundation of China, the Ministry of Science and Technology of China (973 project: No. 2011CBA00105), and the Japan-China Bilateral Joint Research Project by the Japan Society for the Promotion of Science.

\bibliography{references}

\begin{thebibliography}{58}%
\makeatletter
\providecommand \@ifxundefined [1]{%
 \@ifx{#1\undefined}
}%
\providecommand \@ifnum [1]{%
 \ifnum #1\expandafter \@firstoftwo
 \else \expandafter \@secondoftwo
 \fi
}%
\providecommand \@ifx [1]{%
 \ifx #1\expandafter \@firstoftwo
 \else \expandafter \@secondoftwo
 \fi
}%
\providecommand \natexlab [1]{#1}%
\providecommand \enquote  [1]{``#1''}%
\providecommand \bibnamefont  [1]{#1}%
\providecommand \bibfnamefont [1]{#1}%
\providecommand \citenamefont [1]{#1}%
\providecommand \href@noop [0]{\@secondoftwo}%
\providecommand \href [0]{\begingroup \@sanitize@url \@href}%
\providecommand \@href[1]{\@@startlink{#1}\@@href}%
\providecommand \@@href[1]{\endgroup#1\@@endlink}%
\providecommand \@sanitize@url [0]{\catcode `\\12\catcode `\$12\catcode
  `\&12\catcode `\#12\catcode `\^12\catcode `\_12\catcode `\%12\relax}%
\providecommand \@@startlink[1]{}%
\providecommand \@@endlink[0]{}%
\providecommand \url  [0]{\begingroup\@sanitize@url \@url }%
\providecommand \@url [1]{\endgroup\@href {#1}{\urlprefix }}%
\providecommand \urlprefix  [0]{URL }%
\providecommand \Eprint [0]{\href }%
\providecommand \doibase [0]{http://dx.doi.org/}%
\providecommand \selectlanguage [0]{\@gobble}%
\providecommand \bibinfo  [0]{\@secondoftwo}%
\providecommand \bibfield  [0]{\@secondoftwo}%
\providecommand \translation [1]{[#1]}%
\providecommand \BibitemOpen [0]{}%
\providecommand \bibitemStop [0]{}%
\providecommand \bibitemNoStop [0]{.\EOS\space}%
\providecommand \EOS [0]{\spacefactor3000\relax}%
\providecommand \BibitemShut  [1]{\csname bibitem#1\endcsname}%
\let\auto@bib@innerbib\@empty
\bibitem [{\citenamefont {Kamihara}\ \emph {et~al.}(2008)\citenamefont
  {Kamihara}, \citenamefont {Watanabe}, \citenamefont {Hirano},\ and\
  \citenamefont {Hosono}}]{KamiharaJACS}%
  \BibitemOpen
  \bibfield  {author} {\bibinfo {author} {\bibfnamefont {Y.}~\bibnamefont
  {Kamihara}}, \bibinfo {author} {\bibfnamefont {T.}~\bibnamefont {Watanabe}},
  \bibinfo {author} {\bibfnamefont {M.}~\bibnamefont {Hirano}}, \ and\ \bibinfo
  {author} {\bibfnamefont {H.}~\bibnamefont {Hosono}},\ }\href@noop {}
  {\bibfield  {journal} {\bibinfo  {journal} {J. Am. Chem. Soc.}\ }\textbf
  {\bibinfo {volume} {130}},\ \bibinfo {pages} {3296} (\bibinfo {year}
  {2008})}\BibitemShut {NoStop}%
\bibitem [{\citenamefont {Stewart}(2011)}]{StewartIBSsreview}%
  \BibitemOpen
  \bibfield  {author} {\bibinfo {author} {\bibfnamefont {G.~R.}\ \bibnamefont
  {Stewart}},\ }\href@noop {} {\bibfield  {journal} {\bibinfo  {journal} {Rev.
  Mod. Phys.}\ }\textbf {\bibinfo {volume} {83}},\ \bibinfo {pages} {1589}
  (\bibinfo {year} {2011})}\BibitemShut {NoStop}%
\bibitem [{\citenamefont {Mazin}\ \emph {et~al.}(2008)\citenamefont {Mazin},
  \citenamefont {Singh}, \citenamefont {Johannes},\ and\ \citenamefont
  {Du}}]{MazinS}%
  \BibitemOpen
  \bibfield  {author} {\bibinfo {author} {\bibfnamefont {I.~I.}\ \bibnamefont
  {Mazin}}, \bibinfo {author} {\bibfnamefont {D.~J.}\ \bibnamefont {Singh}},
  \bibinfo {author} {\bibfnamefont {M.~D.}\ \bibnamefont {Johannes}}, \ and\
  \bibinfo {author} {\bibfnamefont {M.~H.}\ \bibnamefont {Du}},\ }\href@noop {}
  {\bibfield  {journal} {\bibinfo  {journal} {Phys. Rev. Lett.}\ }\textbf
  {\bibinfo {volume} {101}},\ \bibinfo {pages} {057003} (\bibinfo {year}
  {2008})}\BibitemShut {NoStop}%
\bibitem [{\citenamefont {Kontani}\ and\ \citenamefont
  {Onari}(2010)}]{KontaniPRL}%
  \BibitemOpen
  \bibfield  {author} {\bibinfo {author} {\bibfnamefont {H.}~\bibnamefont
  {Kontani}}\ and\ \bibinfo {author} {\bibfnamefont {S.}~\bibnamefont
  {Onari}},\ }\href@noop {} {\bibfield  {journal} {\bibinfo  {journal} {Phys.
  Rev. Lett.}\ }\textbf {\bibinfo {volume} {104}},\ \bibinfo {pages} {157001}
  (\bibinfo {year} {2010})}\BibitemShut {NoStop}%
\bibitem [{\citenamefont {Rullier-Albenque}\ \emph {et~al.}(2012)\citenamefont
  {Rullier-Albenque}, \citenamefont {Colson}, \citenamefont {Forget},\ and\
  \citenamefont {Alloul}}]{RullierPRL}%
  \BibitemOpen
  \bibfield  {author} {\bibinfo {author} {\bibfnamefont {F.}~\bibnamefont
  {Rullier-Albenque}}, \bibinfo {author} {\bibfnamefont {D.}~\bibnamefont
  {Colson}}, \bibinfo {author} {\bibfnamefont {A.}~\bibnamefont {Forget}}, \
  and\ \bibinfo {author} {\bibfnamefont {H.}~\bibnamefont {Alloul}},\
  }\href@noop {} {\bibfield  {journal} {\bibinfo  {journal} {Phys. Rev. Lett.}\
  }\textbf {\bibinfo {volume} {109}},\ \bibinfo {pages} {187005} (\bibinfo
  {year} {2012})}\BibitemShut {NoStop}%
\bibitem [{\citenamefont {Ohgushi}\ and\ \citenamefont
  {Kiuchi}(2012)}]{OhgushiPRB}%
  \BibitemOpen
  \bibfield  {author} {\bibinfo {author} {\bibfnamefont {K.}~\bibnamefont
  {Ohgushi}}\ and\ \bibinfo {author} {\bibfnamefont {Y.}~\bibnamefont
  {Kiuchi}},\ }\href@noop {} {\bibfield  {journal} {\bibinfo  {journal} {Phys.
  Rev. B}\ }\textbf {\bibinfo {volume} {85}},\ \bibinfo {pages} {064522}
  (\bibinfo {year} {2012})}\BibitemShut {NoStop}%
\bibitem [{\citenamefont {Cheng}\ \emph {et~al.}(2008)\citenamefont {Cheng},
  \citenamefont {Yang}, \citenamefont {Jia}, \citenamefont {Fang},
  \citenamefont {Zhu}, \citenamefont {Mu},\ and\ \citenamefont
  {Wen}}]{ChengPRBNd1111}%
  \BibitemOpen
  \bibfield  {author} {\bibinfo {author} {\bibfnamefont {P.}~\bibnamefont
  {Cheng}}, \bibinfo {author} {\bibfnamefont {H.}~\bibnamefont {Yang}},
  \bibinfo {author} {\bibfnamefont {Y.}~\bibnamefont {Jia}}, \bibinfo {author}
  {\bibfnamefont {L.}~\bibnamefont {Fang}}, \bibinfo {author} {\bibfnamefont
  {X.~Y.}\ \bibnamefont {Zhu}}, \bibinfo {author} {\bibfnamefont
  {G.}~\bibnamefont {Mu}}, \ and\ \bibinfo {author} {\bibfnamefont {H.~H.}\
  \bibnamefont {Wen}},\ }\href@noop {} {\bibfield  {journal} {\bibinfo
  {journal} {Phys. Rev. B}\ }\textbf {\bibinfo {volume} {78}},\ \bibinfo
  {pages} {134508} (\bibinfo {year} {2008})}\BibitemShut {NoStop}%
\bibitem [{\citenamefont {Sun}\ \emph {et~al.}(2011)\citenamefont {Sun},
  \citenamefont {Ding}, \citenamefont {Zheng}, \citenamefont {Shi},\ and\
  \citenamefont {Ren}}]{SunJAP}%
  \BibitemOpen
  \bibfield  {author} {\bibinfo {author} {\bibfnamefont {Y.}~\bibnamefont
  {Sun}}, \bibinfo {author} {\bibfnamefont {Y.}~\bibnamefont {Ding}}, \bibinfo
  {author} {\bibfnamefont {B.~C.}\ \bibnamefont {Zheng}}, \bibinfo {author}
  {\bibfnamefont {Z.~X.}\ \bibnamefont {Shi}}, \ and\ \bibinfo {author}
  {\bibfnamefont {Z.~A.}\ \bibnamefont {Ren}},\ }\href@noop {} {\bibfield
  {journal} {\bibinfo  {journal} {J. Appl. Phys.}\ }\textbf {\bibinfo {volume}
  {109}},\ \bibinfo {pages} {083914} (\bibinfo {year} {2011})}\BibitemShut
  {NoStop}%
\bibitem [{\citenamefont {Richard}\ \emph {et~al.}(2010)\citenamefont
  {Richard}, \citenamefont {Nakayama}, \citenamefont {Sato}, \citenamefont
  {Neupane}, \citenamefont {Xu}, \citenamefont {Bowen}, \citenamefont {Chen},
  \citenamefont {Luo}, \citenamefont {Wang}, \citenamefont {Dai}, \citenamefont
  {Fang}, \citenamefont {Ding},\ and\ \citenamefont {Takahashi}}]{RichardPRL}%
  \BibitemOpen
  \bibfield  {author} {\bibinfo {author} {\bibfnamefont {P.}~\bibnamefont
  {Richard}}, \bibinfo {author} {\bibfnamefont {K.}~\bibnamefont {Nakayama}},
  \bibinfo {author} {\bibfnamefont {T.}~\bibnamefont {Sato}}, \bibinfo {author}
  {\bibfnamefont {M.}~\bibnamefont {Neupane}}, \bibinfo {author} {\bibfnamefont
  {Y.~M.}\ \bibnamefont {Xu}}, \bibinfo {author} {\bibfnamefont {J.~H.}\
  \bibnamefont {Bowen}}, \bibinfo {author} {\bibfnamefont {G.~F.}\ \bibnamefont
  {Chen}}, \bibinfo {author} {\bibfnamefont {J.~L.}\ \bibnamefont {Luo}},
  \bibinfo {author} {\bibfnamefont {N.~L.}\ \bibnamefont {Wang}}, \bibinfo
  {author} {\bibfnamefont {X.}~\bibnamefont {Dai}}, \bibinfo {author}
  {\bibfnamefont {Z.}~\bibnamefont {Fang}}, \bibinfo {author} {\bibfnamefont
  {H.}~\bibnamefont {Ding}}, \ and\ \bibinfo {author} {\bibfnamefont
  {T.}~\bibnamefont {Takahashi}},\ }\href@noop {} {\bibfield  {journal}
  {\bibinfo  {journal} {Phys. Rev. Lett.}\ }\textbf {\bibinfo {volume} {104}},\
  \bibinfo {pages} {137001} (\bibinfo {year} {2010})}\BibitemShut {NoStop}%
\bibitem [{\citenamefont {Huynh}\ \emph {et~al.}(2011)\citenamefont {Huynh},
  \citenamefont {Tanabe},\ and\ \citenamefont {Tanigaki}}]{HuynhPRL}%
  \BibitemOpen
  \bibfield  {author} {\bibinfo {author} {\bibfnamefont {K.~K.}\ \bibnamefont
  {Huynh}}, \bibinfo {author} {\bibfnamefont {Y.}~\bibnamefont {Tanabe}}, \
  and\ \bibinfo {author} {\bibfnamefont {K.}~\bibnamefont {Tanigaki}},\
  }\href@noop {} {\bibfield  {journal} {\bibinfo  {journal} {Phys. Rev. Lett.}\
  }\textbf {\bibinfo {volume} {106}},\ \bibinfo {pages} {217004} (\bibinfo
  {year} {2011})}\BibitemShut {NoStop}%
\bibitem [{\citenamefont {Pallecchi}\ \emph {et~al.}(2011)\citenamefont
  {Pallecchi}, \citenamefont {Bernardini}, \citenamefont {Tropeano},
  \citenamefont {Palenzona}, \citenamefont {Martinelli}, \citenamefont
  {Ferdeghini}, \citenamefont {Vignolo}, \citenamefont {Massidda},\ and\
  \citenamefont {Putti}}]{PallecchiPRB}%
  \BibitemOpen
  \bibfield  {author} {\bibinfo {author} {\bibfnamefont {I.}~\bibnamefont
  {Pallecchi}}, \bibinfo {author} {\bibfnamefont {F.}~\bibnamefont
  {Bernardini}}, \bibinfo {author} {\bibfnamefont {M.}~\bibnamefont
  {Tropeano}}, \bibinfo {author} {\bibfnamefont {A.}~\bibnamefont {Palenzona}},
  \bibinfo {author} {\bibfnamefont {A.}~\bibnamefont {Martinelli}}, \bibinfo
  {author} {\bibfnamefont {C.}~\bibnamefont {Ferdeghini}}, \bibinfo {author}
  {\bibfnamefont {M.}~\bibnamefont {Vignolo}}, \bibinfo {author} {\bibfnamefont
  {S.}~\bibnamefont {Massidda}}, \ and\ \bibinfo {author} {\bibfnamefont
  {M.}~\bibnamefont {Putti}},\ }\href@noop {} {\bibfield  {journal} {\bibinfo
  {journal} {Phys. Rev. B}\ }\textbf {\bibinfo {volume} {84}},\ \bibinfo
  {pages} {134524} (\bibinfo {year} {2011})}\BibitemShut {NoStop}%
\bibitem [{\citenamefont {Chong}\ \emph {et~al.}(2013)\citenamefont {Chong},
  \citenamefont {Williams}, \citenamefont {Kennedy}, \citenamefont {Fang},
  \citenamefont {Tallon},\ and\ \citenamefont {Kadowaki}}]{ChongEPL}%
  \BibitemOpen
  \bibfield  {author} {\bibinfo {author} {\bibfnamefont {S.~V.}\ \bibnamefont
  {Chong}}, \bibinfo {author} {\bibfnamefont {G.~V.~M.}\ \bibnamefont
  {Williams}}, \bibinfo {author} {\bibfnamefont {J.}~\bibnamefont {Kennedy}},
  \bibinfo {author} {\bibfnamefont {F.}~\bibnamefont {Fang}}, \bibinfo {author}
  {\bibfnamefont {J.~L.}\ \bibnamefont {Tallon}}, \ and\ \bibinfo {author}
  {\bibfnamefont {K.}~\bibnamefont {Kadowaki}},\ }\href@noop {} {\bibfield
  {journal} {\bibinfo  {journal} {Europhys. Lett.}\ }\textbf {\bibinfo {volume}
  {104}},\ \bibinfo {pages} {17002} (\bibinfo {year} {2013})}\BibitemShut
  {NoStop}%
\bibitem [{\citenamefont {Bhoi}\ \emph {et~al.}(2011)\citenamefont {Bhoi},
  \citenamefont {Mandal}, \citenamefont {Choudhury}, \citenamefont {Pandya},\
  and\ \citenamefont {Ganesan}}]{BhoiPrFeAsO}%
  \BibitemOpen
  \bibfield  {author} {\bibinfo {author} {\bibfnamefont {D.}~\bibnamefont
  {Bhoi}}, \bibinfo {author} {\bibfnamefont {P.}~\bibnamefont {Mandal}},
  \bibinfo {author} {\bibfnamefont {P.}~\bibnamefont {Choudhury}}, \bibinfo
  {author} {\bibfnamefont {S.}~\bibnamefont {Pandya}}, \ and\ \bibinfo {author}
  {\bibfnamefont {V.}~\bibnamefont {Ganesan}},\ }\href@noop {} {\bibfield
  {journal} {\bibinfo  {journal} {Appl. Phys. Lett.}\ }\textbf {\bibinfo
  {volume} {98}},\ \bibinfo {pages} {172105} (\bibinfo {year}
  {2011})}\BibitemShut {NoStop}%
\bibitem [{\citenamefont {Subedi}\ \emph {et~al.}(2008)\citenamefont {Subedi},
  \citenamefont {Zhang}, \citenamefont {Singh},\ and\ \citenamefont
  {Du}}]{SubediPRB}%
  \BibitemOpen
  \bibfield  {author} {\bibinfo {author} {\bibfnamefont {A.}~\bibnamefont
  {Subedi}}, \bibinfo {author} {\bibfnamefont {L.}~\bibnamefont {Zhang}},
  \bibinfo {author} {\bibfnamefont {D.~J.}\ \bibnamefont {Singh}}, \ and\
  \bibinfo {author} {\bibfnamefont {M.~H.}\ \bibnamefont {Du}},\ }\href@noop {}
  {\bibfield  {journal} {\bibinfo  {journal} {Phys. Rev. B}\ }\textbf {\bibinfo
  {volume} {78}},\ \bibinfo {pages} {134514} (\bibinfo {year}
  {2008})}\BibitemShut {NoStop}%
\bibitem [{\citenamefont {Xia}\ \emph {et~al.}(2009)\citenamefont {Xia},
  \citenamefont {Qian}, \citenamefont {Wray}, \citenamefont {Hsieh},
  \citenamefont {Chen}, \citenamefont {Luo}, \citenamefont {Wang},\ and\
  \citenamefont {Hasan}}]{XiaPRL}%
  \BibitemOpen
  \bibfield  {author} {\bibinfo {author} {\bibfnamefont {Y.}~\bibnamefont
  {Xia}}, \bibinfo {author} {\bibfnamefont {D.}~\bibnamefont {Qian}}, \bibinfo
  {author} {\bibfnamefont {L.}~\bibnamefont {Wray}}, \bibinfo {author}
  {\bibfnamefont {D.}~\bibnamefont {Hsieh}}, \bibinfo {author} {\bibfnamefont
  {G.~F.}\ \bibnamefont {Chen}}, \bibinfo {author} {\bibfnamefont {J.~L.}\
  \bibnamefont {Luo}}, \bibinfo {author} {\bibfnamefont {N.~L.}\ \bibnamefont
  {Wang}}, \ and\ \bibinfo {author} {\bibfnamefont {M.~Z.}\ \bibnamefont
  {Hasan}},\ }\href@noop {} {\bibfield  {journal} {\bibinfo  {journal} {Phys.
  Rev. Lett.}\ }\textbf {\bibinfo {volume} {103}},\ \bibinfo {pages} {037002}
  (\bibinfo {year} {2009})}\BibitemShut {NoStop}%
\bibitem [{\citenamefont {Chen}\ \emph {et~al.}(2010)\citenamefont {Chen},
  \citenamefont {Zhou}, \citenamefont {Zhang}, \citenamefont {Wei},
  \citenamefont {Ou}, \citenamefont {Zhao}, \citenamefont {He}, \citenamefont
  {Ge}, \citenamefont {Arita}, \citenamefont {Shimada}, \citenamefont
  {Namatame}, \citenamefont {Taniguchi}, \citenamefont {Lu}, \citenamefont
  {Hu}, \citenamefont {Cui},\ and\ \citenamefont {Feng}}]{ChenPRB11}%
  \BibitemOpen
  \bibfield  {author} {\bibinfo {author} {\bibfnamefont {F.}~\bibnamefont
  {Chen}}, \bibinfo {author} {\bibfnamefont {B.}~\bibnamefont {Zhou}}, \bibinfo
  {author} {\bibfnamefont {Y.}~\bibnamefont {Zhang}}, \bibinfo {author}
  {\bibfnamefont {J.}~\bibnamefont {Wei}}, \bibinfo {author} {\bibfnamefont
  {H.~W.}\ \bibnamefont {Ou}}, \bibinfo {author} {\bibfnamefont {J.~F.}\
  \bibnamefont {Zhao}}, \bibinfo {author} {\bibfnamefont {C.}~\bibnamefont
  {He}}, \bibinfo {author} {\bibfnamefont {Q.~Q.}\ \bibnamefont {Ge}}, \bibinfo
  {author} {\bibfnamefont {M.}~\bibnamefont {Arita}}, \bibinfo {author}
  {\bibfnamefont {K.}~\bibnamefont {Shimada}}, \bibinfo {author} {\bibfnamefont
  {H.}~\bibnamefont {Namatame}}, \bibinfo {author} {\bibfnamefont
  {M.}~\bibnamefont {Taniguchi}}, \bibinfo {author} {\bibfnamefont {Z.~Y.}\
  \bibnamefont {Lu}}, \bibinfo {author} {\bibfnamefont {J.}~\bibnamefont {Hu}},
  \bibinfo {author} {\bibfnamefont {X.~Y.}\ \bibnamefont {Cui}}, \ and\
  \bibinfo {author} {\bibfnamefont {D.~L.}\ \bibnamefont {Feng}},\ }\href@noop
  {} {\bibfield  {journal} {\bibinfo  {journal} {Phys. Rev. B}\ }\textbf
  {\bibinfo {volume} {81}},\ \bibinfo {pages} {014526} (\bibinfo {year}
  {2010})}\BibitemShut {NoStop}%
\bibitem [{\citenamefont {Chang}\ \emph {et~al.}(2012)\citenamefont {Chang},
  \citenamefont {Luo}, \citenamefont {Wu}, \citenamefont {Hsu}, \citenamefont
  {Huang}, \citenamefont {Wu}, \citenamefont {Wu},\ and\ \citenamefont
  {Wang}}]{ChangSUST}%
  \BibitemOpen
  \bibfield  {author} {\bibinfo {author} {\bibfnamefont {H.~H.}\ \bibnamefont
  {Chang}}, \bibinfo {author} {\bibfnamefont {J.~Y.}\ \bibnamefont {Luo}},
  \bibinfo {author} {\bibfnamefont {C.~T.}\ \bibnamefont {Wu}}, \bibinfo
  {author} {\bibfnamefont {F.~C.}\ \bibnamefont {Hsu}}, \bibinfo {author}
  {\bibfnamefont {T.~W.}\ \bibnamefont {Huang}}, \bibinfo {author}
  {\bibfnamefont {P.~M.}\ \bibnamefont {Wu}}, \bibinfo {author} {\bibfnamefont
  {M.~K.}\ \bibnamefont {Wu}}, \ and\ \bibinfo {author} {\bibfnamefont {M.~J.}\
  \bibnamefont {Wang}},\ }\href@noop {} {\bibfield  {journal} {\bibinfo
  {journal} {Supercond. Sci. Technol.}\ }\textbf {\bibinfo {volume} {25}},\
  \bibinfo {pages} {035004} (\bibinfo {year} {2012})}\BibitemShut {NoStop}%
\bibitem [{\citenamefont {Hu}\ \emph {et~al.}(2013)\citenamefont {Hu},
  \citenamefont {Liu}, \citenamefont {Qian},\ and\ \citenamefont
  {Mao}}]{HuPRB}%
  \BibitemOpen
  \bibfield  {author} {\bibinfo {author} {\bibfnamefont {J.}~\bibnamefont
  {Hu}}, \bibinfo {author} {\bibfnamefont {T.~J.}\ \bibnamefont {Liu}},
  \bibinfo {author} {\bibfnamefont {B.}~\bibnamefont {Qian}}, \ and\ \bibinfo
  {author} {\bibfnamefont {Z.~Q.}\ \bibnamefont {Mao}},\ }\href@noop {}
  {\bibfield  {journal} {\bibinfo  {journal} {Phys. Rev. B}\ }\textbf {\bibinfo
  {volume} {88}},\ \bibinfo {pages} {094505} (\bibinfo {year}
  {2013})}\BibitemShut {NoStop}%
\bibitem [{\citenamefont {Bao}\ \emph {et~al.}(2009)\citenamefont {Bao},
  \citenamefont {Qiu}, \citenamefont {Huang}, \citenamefont {Green},
  \citenamefont {Zajdel}, \citenamefont {Fitzsimmons}, \citenamefont
  {Zhernenkov}, \citenamefont {Chang}, \citenamefont {Fang}, \citenamefont
  {Qian}, \citenamefont {Vehstedt}, \citenamefont {Yang}, \citenamefont {Pham},
  \citenamefont {Spinu},\ and\ \citenamefont {Mao}}]{BaoWeiPRL}%
  \BibitemOpen
  \bibfield  {author} {\bibinfo {author} {\bibfnamefont {W.}~\bibnamefont
  {Bao}}, \bibinfo {author} {\bibfnamefont {Y.}~\bibnamefont {Qiu}}, \bibinfo
  {author} {\bibfnamefont {Q.}~\bibnamefont {Huang}}, \bibinfo {author}
  {\bibfnamefont {M.~A.}\ \bibnamefont {Green}}, \bibinfo {author}
  {\bibfnamefont {P.}~\bibnamefont {Zajdel}}, \bibinfo {author} {\bibfnamefont
  {M.~R.}\ \bibnamefont {Fitzsimmons}}, \bibinfo {author} {\bibfnamefont
  {M.}~\bibnamefont {Zhernenkov}}, \bibinfo {author} {\bibfnamefont
  {S.}~\bibnamefont {Chang}}, \bibinfo {author} {\bibfnamefont
  {M.}~\bibnamefont {Fang}}, \bibinfo {author} {\bibfnamefont {B.}~\bibnamefont
  {Qian}}, \bibinfo {author} {\bibfnamefont {E.~K.}\ \bibnamefont {Vehstedt}},
  \bibinfo {author} {\bibfnamefont {J.}~\bibnamefont {Yang}}, \bibinfo {author}
  {\bibfnamefont {H.~M.}\ \bibnamefont {Pham}}, \bibinfo {author}
  {\bibfnamefont {L.}~\bibnamefont {Spinu}}, \ and\ \bibinfo {author}
  {\bibfnamefont {Z.~Q.}\ \bibnamefont {Mao}},\ }\href@noop {} {\bibfield
  {journal} {\bibinfo  {journal} {Phys. Rev. Lett.}\ }\textbf {\bibinfo
  {volume} {102}},\ \bibinfo {pages} {247001} (\bibinfo {year}
  {2009})}\BibitemShut {NoStop}%
\bibitem [{\citenamefont {Bendele}\ \emph {et~al.}(2010)\citenamefont
  {Bendele}, \citenamefont {Babkevich}, \citenamefont {Katrych}, \citenamefont
  {Gvasaliya}, \citenamefont {Pomjakushina}, \citenamefont {Conder},
  \citenamefont {Roessli}, \citenamefont {Boothroyd}, \citenamefont
  {Khasanov},\ and\ \citenamefont {Keller}}]{BendelePRB}%
  \BibitemOpen
  \bibfield  {author} {\bibinfo {author} {\bibfnamefont {M.}~\bibnamefont
  {Bendele}}, \bibinfo {author} {\bibfnamefont {P.}~\bibnamefont {Babkevich}},
  \bibinfo {author} {\bibfnamefont {S.}~\bibnamefont {Katrych}}, \bibinfo
  {author} {\bibfnamefont {S.~N.}\ \bibnamefont {Gvasaliya}}, \bibinfo {author}
  {\bibfnamefont {E.}~\bibnamefont {Pomjakushina}}, \bibinfo {author}
  {\bibfnamefont {K.}~\bibnamefont {Conder}}, \bibinfo {author} {\bibfnamefont
  {B.}~\bibnamefont {Roessli}}, \bibinfo {author} {\bibfnamefont {A.~T.}\
  \bibnamefont {Boothroyd}}, \bibinfo {author} {\bibfnamefont {R.}~\bibnamefont
  {Khasanov}}, \ and\ \bibinfo {author} {\bibfnamefont {H.}~\bibnamefont
  {Keller}},\ }\href@noop {} {\bibfield  {journal} {\bibinfo  {journal} {Phys.
  Rev. B}\ }\textbf {\bibinfo {volume} {82}},\ \bibinfo {pages} {212504}
  (\bibinfo {year} {2010})}\BibitemShut {NoStop}%
\bibitem [{\citenamefont {Zhang}\ \emph {et~al.}(2009)\citenamefont {Zhang},
  \citenamefont {Singh},\ and\ \citenamefont {Du}}]{ZhangPRB}%
  \BibitemOpen
  \bibfield  {author} {\bibinfo {author} {\bibfnamefont {L.}~\bibnamefont
  {Zhang}}, \bibinfo {author} {\bibfnamefont {D.~J.}\ \bibnamefont {Singh}}, \
  and\ \bibinfo {author} {\bibfnamefont {M.~H.}\ \bibnamefont {Du}},\
  }\href@noop {} {\bibfield  {journal} {\bibinfo  {journal} {Phys. Rev. B}\
  }\textbf {\bibinfo {volume} {79}},\ \bibinfo {pages} {012506} (\bibinfo
  {year} {2009})}\BibitemShut {NoStop}%
\bibitem [{\citenamefont {Thampy}\ \emph {et~al.}(2012)\citenamefont {Thampy},
  \citenamefont {Kang}, \citenamefont {Rodriguez-Rivera}, \citenamefont {Bao},
  \citenamefont {Savici}, \citenamefont {Hu}, \citenamefont {Liu},
  \citenamefont {Qian}, \citenamefont {Fobes}, \citenamefont {Mao},
  \citenamefont {Fu}, \citenamefont {Chen}, \citenamefont {Ye}, \citenamefont
  {Erwin}, \citenamefont {Gentile}, \citenamefont {Tesanovic},\ and\
  \citenamefont {Broholm}}]{ThampyPRL}%
  \BibitemOpen
  \bibfield  {author} {\bibinfo {author} {\bibfnamefont {V.}~\bibnamefont
  {Thampy}}, \bibinfo {author} {\bibfnamefont {J.}~\bibnamefont {Kang}},
  \bibinfo {author} {\bibfnamefont {J.~A.}\ \bibnamefont {Rodriguez-Rivera}},
  \bibinfo {author} {\bibfnamefont {W.}~\bibnamefont {Bao}}, \bibinfo {author}
  {\bibfnamefont {A.~T.}\ \bibnamefont {Savici}}, \bibinfo {author}
  {\bibfnamefont {J.}~\bibnamefont {Hu}}, \bibinfo {author} {\bibfnamefont
  {T.~J.}\ \bibnamefont {Liu}}, \bibinfo {author} {\bibfnamefont
  {B.}~\bibnamefont {Qian}}, \bibinfo {author} {\bibfnamefont {D.}~\bibnamefont
  {Fobes}}, \bibinfo {author} {\bibfnamefont {Z.~Q.}\ \bibnamefont {Mao}},
  \bibinfo {author} {\bibfnamefont {C.~B.}\ \bibnamefont {Fu}}, \bibinfo
  {author} {\bibfnamefont {W.~C.}\ \bibnamefont {Chen}}, \bibinfo {author}
  {\bibfnamefont {Q.}~\bibnamefont {Ye}}, \bibinfo {author} {\bibfnamefont
  {R.~W.}\ \bibnamefont {Erwin}}, \bibinfo {author} {\bibfnamefont {T.~R.}\
  \bibnamefont {Gentile}}, \bibinfo {author} {\bibfnamefont {Z.}~\bibnamefont
  {Tesanovic}}, \ and\ \bibinfo {author} {\bibfnamefont {C.}~\bibnamefont
  {Broholm}},\ }\href@noop {} {\bibfield  {journal} {\bibinfo  {journal} {Phys.
  Rev. Lett.}\ }\textbf {\bibinfo {volume} {108}},\ \bibinfo {pages} {107002}
  (\bibinfo {year} {2012})}\BibitemShut {NoStop}%
\bibitem [{\citenamefont {Liu}\ \emph {et~al.}(2009)\citenamefont {Liu},
  \citenamefont {Ke}, \citenamefont {Qian}, \citenamefont {Hu}, \citenamefont
  {Fobes}, \citenamefont {Vehstedt}, \citenamefont {Pham}, \citenamefont
  {Yang}, \citenamefont {Fang}, \citenamefont {Spinu}, \citenamefont
  {Schiffer}, \citenamefont {Liu},\ and\ \citenamefont {Mao}}]{LiuPRB}%
  \BibitemOpen
  \bibfield  {author} {\bibinfo {author} {\bibfnamefont {T.~J.}\ \bibnamefont
  {Liu}}, \bibinfo {author} {\bibfnamefont {X.}~\bibnamefont {Ke}}, \bibinfo
  {author} {\bibfnamefont {B.}~\bibnamefont {Qian}}, \bibinfo {author}
  {\bibfnamefont {J.}~\bibnamefont {Hu}}, \bibinfo {author} {\bibfnamefont
  {D.}~\bibnamefont {Fobes}}, \bibinfo {author} {\bibfnamefont {E.~K.}\
  \bibnamefont {Vehstedt}}, \bibinfo {author} {\bibfnamefont {H.}~\bibnamefont
  {Pham}}, \bibinfo {author} {\bibfnamefont {J.~H.}\ \bibnamefont {Yang}},
  \bibinfo {author} {\bibfnamefont {M.~H.}\ \bibnamefont {Fang}}, \bibinfo
  {author} {\bibfnamefont {L.}~\bibnamefont {Spinu}}, \bibinfo {author}
  {\bibfnamefont {P.}~\bibnamefont {Schiffer}}, \bibinfo {author}
  {\bibfnamefont {Y.}~\bibnamefont {Liu}}, \ and\ \bibinfo {author}
  {\bibfnamefont {Z.~Q.}\ \bibnamefont {Mao}},\ }\href@noop {} {\bibfield
  {journal} {\bibinfo  {journal} {Phys. Rev. B}\ }\textbf {\bibinfo {volume}
  {80}},\ \bibinfo {pages} {174509} (\bibinfo {year} {2009})}\BibitemShut
  {NoStop}%
\bibitem [{\citenamefont {Sun}\ \emph {et~al.}(2013{\natexlab{a}})\citenamefont
  {Sun}, \citenamefont {Taen}, \citenamefont {Tsuchiya}, \citenamefont {Ding},
  \citenamefont {Pyon}, \citenamefont {Shi},\ and\ \citenamefont
  {Tamegai}}]{SunAPRE}%
  \BibitemOpen
  \bibfield  {author} {\bibinfo {author} {\bibfnamefont {Y.}~\bibnamefont
  {Sun}}, \bibinfo {author} {\bibfnamefont {T.}~\bibnamefont {Taen}}, \bibinfo
  {author} {\bibfnamefont {Y.}~\bibnamefont {Tsuchiya}}, \bibinfo {author}
  {\bibfnamefont {Q.}~\bibnamefont {Ding}}, \bibinfo {author} {\bibfnamefont
  {S.}~\bibnamefont {Pyon}}, \bibinfo {author} {\bibfnamefont {Z.~X.}\
  \bibnamefont {Shi}}, \ and\ \bibinfo {author} {\bibfnamefont
  {T.}~\bibnamefont {Tamegai}},\ }\href@noop {} {\bibfield  {journal} {\bibinfo
   {journal} {Appl. Phys. Express}\ }\textbf {\bibinfo {volume} {6}},\ \bibinfo
  {pages} {043101} (\bibinfo {year} {2013}{\natexlab{a}})}\BibitemShut
  {NoStop}%
\bibitem [{\citenamefont {Sun}\ \emph {et~al.}(2013{\natexlab{b}})\citenamefont
  {Sun}, \citenamefont {Taen}, \citenamefont {Tsuchiya}, \citenamefont {Shi},\
  and\ \citenamefont {Tamegai}}]{SunSUST}%
  \BibitemOpen
  \bibfield  {author} {\bibinfo {author} {\bibfnamefont {Y.}~\bibnamefont
  {Sun}}, \bibinfo {author} {\bibfnamefont {T.}~\bibnamefont {Taen}}, \bibinfo
  {author} {\bibfnamefont {Y.}~\bibnamefont {Tsuchiya}}, \bibinfo {author}
  {\bibfnamefont {Z.~X.}\ \bibnamefont {Shi}}, \ and\ \bibinfo {author}
  {\bibfnamefont {T.}~\bibnamefont {Tamegai}},\ }\href@noop {} {\bibfield
  {journal} {\bibinfo  {journal} {Supercond. Sci. Technol.}\ }\textbf {\bibinfo
  {volume} {26}},\ \bibinfo {pages} {015015} (\bibinfo {year}
  {2013}{\natexlab{b}})}\BibitemShut {NoStop}%
\bibitem [{\citenamefont {Sun}\ \emph {et~al.}(2013{\natexlab{c}})\citenamefont
  {Sun}, \citenamefont {Tsuchiya}, \citenamefont {Yamada}, \citenamefont
  {Taen}, \citenamefont {Pyon}, \citenamefont {Shi},\ and\ \citenamefont
  {Tamegai}}]{Sunjpsj}%
  \BibitemOpen
  \bibfield  {author} {\bibinfo {author} {\bibfnamefont {Y.}~\bibnamefont
  {Sun}}, \bibinfo {author} {\bibfnamefont {Y.}~\bibnamefont {Tsuchiya}},
  \bibinfo {author} {\bibfnamefont {T.}~\bibnamefont {Yamada}}, \bibinfo
  {author} {\bibfnamefont {T.}~\bibnamefont {Taen}}, \bibinfo {author}
  {\bibfnamefont {S.}~\bibnamefont {Pyon}}, \bibinfo {author} {\bibfnamefont
  {Z.~X.}\ \bibnamefont {Shi}}, \ and\ \bibinfo {author} {\bibfnamefont
  {T.}~\bibnamefont {Tamegai}},\ }\href@noop {} {\bibfield  {journal} {\bibinfo
   {journal} {J. Phys. Soc. Jpn.}\ }\textbf {\bibinfo {volume} {82}},\ \bibinfo
  {pages} {093705} (\bibinfo {year} {2013}{\natexlab{c}})}\BibitemShut
  {NoStop}%
\bibitem [{\citenamefont {Sun}\ \emph {et~al.}(2013{\natexlab{d}})\citenamefont
  {Sun}, \citenamefont {Tsuchiya}, \citenamefont {Yamada}, \citenamefont
  {Taen}, \citenamefont {Pyon}, \citenamefont {Shi},\ and\ \citenamefont
  {Tamegai}}]{SunJPSJshort}%
  \BibitemOpen
  \bibfield  {author} {\bibinfo {author} {\bibfnamefont {Y.}~\bibnamefont
  {Sun}}, \bibinfo {author} {\bibfnamefont {Y.}~\bibnamefont {Tsuchiya}},
  \bibinfo {author} {\bibfnamefont {T.}~\bibnamefont {Yamada}}, \bibinfo
  {author} {\bibfnamefont {T.}~\bibnamefont {Taen}}, \bibinfo {author}
  {\bibfnamefont {S.}~\bibnamefont {Pyon}}, \bibinfo {author} {\bibfnamefont
  {Z.~X.}\ \bibnamefont {Shi}}, \ and\ \bibinfo {author} {\bibfnamefont
  {T.}~\bibnamefont {Tamegai}},\ }\href@noop {} {\bibfield  {journal} {\bibinfo
   {journal} {J. Phys. Soc. Jpn.}\ }\textbf {\bibinfo {volume} {82}},\ \bibinfo
  {pages} {115002} (\bibinfo {year} {2013}{\natexlab{d}})}\BibitemShut
  {NoStop}%
\bibitem [{\citenamefont {Sun}\ \emph {et~al.}(2013{\natexlab{e}})\citenamefont
  {Sun}, \citenamefont {Taen}, \citenamefont {Tsuchiya}, \citenamefont {Pyon},
  \citenamefont {Shi},\ and\ \citenamefont {Tamegai}}]{SunEPL}%
  \BibitemOpen
  \bibfield  {author} {\bibinfo {author} {\bibfnamefont {Y.}~\bibnamefont
  {Sun}}, \bibinfo {author} {\bibfnamefont {T.}~\bibnamefont {Taen}}, \bibinfo
  {author} {\bibfnamefont {Y.}~\bibnamefont {Tsuchiya}}, \bibinfo {author}
  {\bibfnamefont {S.}~\bibnamefont {Pyon}}, \bibinfo {author} {\bibfnamefont
  {Z.~X.}\ \bibnamefont {Shi}}, \ and\ \bibinfo {author} {\bibfnamefont
  {T.}~\bibnamefont {Tamegai}},\ }\href@noop {} {\bibfield  {journal} {\bibinfo
   {journal} {Europhys. Lett.}\ }\textbf {\bibinfo {volume} {103}},\ \bibinfo
  {pages} {57013} (\bibinfo {year} {2013}{\natexlab{e}})}\BibitemShut {NoStop}%
\bibitem [{\citenamefont {Sun}\ \emph {et~al.}(2014)\citenamefont {Sun},
  \citenamefont {Tsuchiya}, \citenamefont {Taen}, \citenamefont {Yamada},
  \citenamefont {Pyon}, \citenamefont {Sugimoto}, \citenamefont {Ekino},
  \citenamefont {Shi},\ and\ \citenamefont {Tamegai}}]{SunSciRep}%
  \BibitemOpen
  \bibfield  {author} {\bibinfo {author} {\bibfnamefont {Y.}~\bibnamefont
  {Sun}}, \bibinfo {author} {\bibfnamefont {Y.}~\bibnamefont {Tsuchiya}},
  \bibinfo {author} {\bibfnamefont {T.}~\bibnamefont {Taen}}, \bibinfo {author}
  {\bibfnamefont {T.}~\bibnamefont {Yamada}}, \bibinfo {author} {\bibfnamefont
  {S.}~\bibnamefont {Pyon}}, \bibinfo {author} {\bibfnamefont {A.}~\bibnamefont
  {Sugimoto}}, \bibinfo {author} {\bibfnamefont {T.}~\bibnamefont {Ekino}},
  \bibinfo {author} {\bibfnamefont {Z.~X.}\ \bibnamefont {Shi}}, \ and\
  \bibinfo {author} {\bibfnamefont {T.}~\bibnamefont {Tamegai}},\ }\href@noop
  {} {\bibfield  {journal} {\bibinfo  {journal} {Sci. Rep}\ }\textbf {\bibinfo
  {volume} {4}},\ \bibinfo {pages} {4585} (\bibinfo {year} {2014})}\BibitemShut
  {NoStop}%
\bibitem [{\citenamefont {Taen}\ \emph {et~al.}(2009)\citenamefont {Taen},
  \citenamefont {Tsuchiya}, \citenamefont {Nakajima},\ and\ \citenamefont
  {Tamegai}}]{TaenPRB}%
  \BibitemOpen
  \bibfield  {author} {\bibinfo {author} {\bibfnamefont {T.}~\bibnamefont
  {Taen}}, \bibinfo {author} {\bibfnamefont {Y.}~\bibnamefont {Tsuchiya}},
  \bibinfo {author} {\bibfnamefont {Y.}~\bibnamefont {Nakajima}}, \ and\
  \bibinfo {author} {\bibfnamefont {T.}~\bibnamefont {Tamegai}},\ }\href@noop
  {} {\bibfield  {journal} {\bibinfo  {journal} {Phys. Rev. B}\ }\textbf
  {\bibinfo {volume} {80}},\ \bibinfo {pages} {092502} (\bibinfo {year}
  {2009})}\BibitemShut {NoStop}%
\bibitem [{\citenamefont {Dong}\ \emph {et~al.}(2011)\citenamefont {Dong},
  \citenamefont {Wang}, \citenamefont {Li}, \citenamefont {Chen}, \citenamefont
  {Yuan},\ and\ \citenamefont {Fang}}]{DongPRB}%
  \BibitemOpen
  \bibfield  {author} {\bibinfo {author} {\bibfnamefont {C.}~\bibnamefont
  {Dong}}, \bibinfo {author} {\bibfnamefont {H.}~\bibnamefont {Wang}}, \bibinfo
  {author} {\bibfnamefont {Z.}~\bibnamefont {Li}}, \bibinfo {author}
  {\bibfnamefont {J.}~\bibnamefont {Chen}}, \bibinfo {author} {\bibfnamefont
  {H.~Q.}\ \bibnamefont {Yuan}}, \ and\ \bibinfo {author} {\bibfnamefont
  {M.}~\bibnamefont {Fang}},\ }\href@noop {} {\bibfield  {journal} {\bibinfo
  {journal} {Phys. Rev. B}\ }\textbf {\bibinfo {volume} {84}},\ \bibinfo
  {pages} {224506} (\bibinfo {year} {2011})}\BibitemShut {NoStop}%
\bibitem [{\citenamefont {Lei}\ \emph {et~al.}(2012)\citenamefont {Lei},
  \citenamefont {Graf}, \citenamefont {Hu}, \citenamefont {Ryu}, \citenamefont
  {Choi}, \citenamefont {Tozer},\ and\ \citenamefont {Petrovic}}]{LeiPRB}%
  \BibitemOpen
  \bibfield  {author} {\bibinfo {author} {\bibfnamefont {H.}~\bibnamefont
  {Lei}}, \bibinfo {author} {\bibfnamefont {D.}~\bibnamefont {Graf}}, \bibinfo
  {author} {\bibfnamefont {R.}~\bibnamefont {Hu}}, \bibinfo {author}
  {\bibfnamefont {H.}~\bibnamefont {Ryu}}, \bibinfo {author} {\bibfnamefont
  {E.~S.}\ \bibnamefont {Choi}}, \bibinfo {author} {\bibfnamefont {S.~W.}\
  \bibnamefont {Tozer}}, \ and\ \bibinfo {author} {\bibfnamefont
  {C.}~\bibnamefont {Petrovic}},\ }\href@noop {} {\bibfield  {journal}
  {\bibinfo  {journal} {Phys. Rev. B}\ }\textbf {\bibinfo {volume} {85}},\
  \bibinfo {pages} {094515} (\bibinfo {year} {2012})}\BibitemShut {NoStop}%
\bibitem [{\citenamefont {Tsukada}\ \emph {et~al.}(2010)\citenamefont
  {Tsukada}, \citenamefont {Hanawa}, \citenamefont {Komiya}, \citenamefont
  {Akiike}, \citenamefont {Tanaka}, \citenamefont {Imai},\ and\ \citenamefont
  {Maeda}}]{TsukadaPRB}%
  \BibitemOpen
  \bibfield  {author} {\bibinfo {author} {\bibfnamefont {I.}~\bibnamefont
  {Tsukada}}, \bibinfo {author} {\bibfnamefont {M.}~\bibnamefont {Hanawa}},
  \bibinfo {author} {\bibfnamefont {S.}~\bibnamefont {Komiya}}, \bibinfo
  {author} {\bibfnamefont {T.}~\bibnamefont {Akiike}}, \bibinfo {author}
  {\bibfnamefont {R.}~\bibnamefont {Tanaka}}, \bibinfo {author} {\bibfnamefont
  {Y.}~\bibnamefont {Imai}}, \ and\ \bibinfo {author} {\bibfnamefont
  {A.}~\bibnamefont {Maeda}},\ }\href@noop {} {\bibfield  {journal} {\bibinfo
  {journal} {Phys. Rev. B}\ }\textbf {\bibinfo {volume} {81}},\ \bibinfo
  {pages} {054515} (\bibinfo {year} {2010})}\BibitemShut {NoStop}%
\bibitem [{\citenamefont {Smith}(1978)}]{SmithBook}%
  \BibitemOpen
  \bibfield  {author} {\bibinfo {author} {\bibfnamefont {R.~A.}\ \bibnamefont
  {Smith}},\ }\href@noop {} {\emph {\bibinfo {title} {Semiconductors}}}\
  (\bibinfo  {publisher} {Cambridge University Press},\ \bibinfo {year}
  {1978})\BibitemShut {NoStop}%
\bibitem [{\citenamefont {He}\ \emph {et~al.}(2013)\citenamefont {He},
  \citenamefont {He}, \citenamefont {Zhang}, \citenamefont {Zhao},
  \citenamefont {Liu}, \citenamefont {Liu}, \citenamefont {Mou}, \citenamefont
  {Ou}, \citenamefont {Wang}, \citenamefont {Li}, \citenamefont {Wang},
  \citenamefont {Peng}, \citenamefont {Liu}, \citenamefont {Chen},
  \citenamefont {Yu}, \citenamefont {Liu}, \citenamefont {Dong}, \citenamefont
  {Zhang}, \citenamefont {Chen}, \citenamefont {Xu}, \citenamefont {Chen},
  \citenamefont {Ma}, \citenamefont {Xue},\ and\ \citenamefont
  {Zhou}}]{HeNatMat}%
  \BibitemOpen
  \bibfield  {author} {\bibinfo {author} {\bibfnamefont {S.}~\bibnamefont
  {He}}, \bibinfo {author} {\bibfnamefont {J.}~\bibnamefont {He}}, \bibinfo
  {author} {\bibfnamefont {W.}~\bibnamefont {Zhang}}, \bibinfo {author}
  {\bibfnamefont {L.}~\bibnamefont {Zhao}}, \bibinfo {author} {\bibfnamefont
  {D.}~\bibnamefont {Liu}}, \bibinfo {author} {\bibfnamefont {X.}~\bibnamefont
  {Liu}}, \bibinfo {author} {\bibfnamefont {D.}~\bibnamefont {Mou}}, \bibinfo
  {author} {\bibfnamefont {Y.~B.}\ \bibnamefont {Ou}}, \bibinfo {author}
  {\bibfnamefont {Q.~Y.}\ \bibnamefont {Wang}}, \bibinfo {author}
  {\bibfnamefont {Z.}~\bibnamefont {Li}}, \bibinfo {author} {\bibfnamefont
  {L.}~\bibnamefont {Wang}}, \bibinfo {author} {\bibfnamefont {Y.}~\bibnamefont
  {Peng}}, \bibinfo {author} {\bibfnamefont {Y.}~\bibnamefont {Liu}}, \bibinfo
  {author} {\bibfnamefont {C.}~\bibnamefont {Chen}}, \bibinfo {author}
  {\bibfnamefont {L.}~\bibnamefont {Yu}}, \bibinfo {author} {\bibfnamefont
  {G.}~\bibnamefont {Liu}}, \bibinfo {author} {\bibfnamefont {X.}~\bibnamefont
  {Dong}}, \bibinfo {author} {\bibfnamefont {J.}~\bibnamefont {Zhang}},
  \bibinfo {author} {\bibfnamefont {C.}~\bibnamefont {Chen}}, \bibinfo {author}
  {\bibfnamefont {Z.}~\bibnamefont {Xu}}, \bibinfo {author} {\bibfnamefont
  {X.}~\bibnamefont {Chen}}, \bibinfo {author} {\bibfnamefont {X.}~\bibnamefont
  {Ma}}, \bibinfo {author} {\bibfnamefont {Q.}~\bibnamefont {Xue}}, \ and\
  \bibinfo {author} {\bibfnamefont {X.~J.}\ \bibnamefont {Zhou}},\ }\href@noop
  {} {\bibfield  {journal} {\bibinfo  {journal} {Nat. Mater.}\ }\textbf
  {\bibinfo {volume} {12}},\ \bibinfo {pages} {605} (\bibinfo {year}
  {2013})}\BibitemShut {NoStop}%
\bibitem [{\citenamefont {Eom}\ \emph {et~al.}(2012)\citenamefont {Eom},
  \citenamefont {Na}, \citenamefont {Hoch}, \citenamefont {Kremer},\ and\
  \citenamefont {Kim}}]{EomPRB}%
  \BibitemOpen
  \bibfield  {author} {\bibinfo {author} {\bibfnamefont {M.~J.}\ \bibnamefont
  {Eom}}, \bibinfo {author} {\bibfnamefont {S.~W.}\ \bibnamefont {Na}},
  \bibinfo {author} {\bibfnamefont {C.}~\bibnamefont {Hoch}}, \bibinfo {author}
  {\bibfnamefont {R.~K.}\ \bibnamefont {Kremer}}, \ and\ \bibinfo {author}
  {\bibfnamefont {J.~S.}\ \bibnamefont {Kim}},\ }\href@noop {} {\bibfield
  {journal} {\bibinfo  {journal} {Phys. Rev. B}\ }\textbf {\bibinfo {volume}
  {85}},\ \bibinfo {pages} {024536} (\bibinfo {year} {2012})}\BibitemShut
  {NoStop}%
\bibitem [{\citenamefont {Ziman}(2001)}]{ZimanBook}%
  \BibitemOpen
  \bibfield  {author} {\bibinfo {author} {\bibfnamefont {J.~M.}\ \bibnamefont
  {Ziman}},\ }\href@noop {} {\emph {\bibinfo {title} {Electrons and Phonons,
  Classics Series}}}\ (\bibinfo  {publisher} {Cambridge University Press},\
  \bibinfo {year} {2001})\BibitemShut {NoStop}%
\bibitem [{\citenamefont {Harris}\ \emph {et~al.}(1995)\citenamefont {Harris},
  \citenamefont {Yan}, \citenamefont {Matl}, \citenamefont {Ong}, \citenamefont
  {Anderson}, \citenamefont {Kimura},\ and\ \citenamefont
  {Kitazawa}}]{HarrisPRL}%
  \BibitemOpen
  \bibfield  {author} {\bibinfo {author} {\bibfnamefont {J.~M.}\ \bibnamefont
  {Harris}}, \bibinfo {author} {\bibfnamefont {Y.~F.}\ \bibnamefont {Yan}},
  \bibinfo {author} {\bibfnamefont {P.}~\bibnamefont {Matl}}, \bibinfo {author}
  {\bibfnamefont {N.~P.}\ \bibnamefont {Ong}}, \bibinfo {author} {\bibfnamefont
  {P.~W.}\ \bibnamefont {Anderson}}, \bibinfo {author} {\bibfnamefont
  {T.}~\bibnamefont {Kimura}}, \ and\ \bibinfo {author} {\bibfnamefont
  {K.}~\bibnamefont {Kitazawa}},\ }\href@noop {} {\bibfield  {journal}
  {\bibinfo  {journal} {Phys. Rev. Lett.}\ }\textbf {\bibinfo {volume} {75}},\
  \bibinfo {pages} {1391} (\bibinfo {year} {1995})}\BibitemShut {NoStop}%
\bibitem [{\citenamefont {Nakajima}\ \emph {et~al.}(2007)\citenamefont
  {Nakajima}, \citenamefont {Shishido}, \citenamefont {Nakai}, \citenamefont
  {Shibauchi}, \citenamefont {Behnia}, \citenamefont {Izawa}, \citenamefont
  {Hedo}, \citenamefont {Uwatoko}, \citenamefont {Matsumoto}, \citenamefont
  {Settai}, \citenamefont {\={O}nuki}, \citenamefont {Kontani},\ and\
  \citenamefont {Matsuda}}]{NakajimaJPSJ}%
  \BibitemOpen
  \bibfield  {author} {\bibinfo {author} {\bibfnamefont {Y.}~\bibnamefont
  {Nakajima}}, \bibinfo {author} {\bibfnamefont {H.}~\bibnamefont {Shishido}},
  \bibinfo {author} {\bibfnamefont {H.}~\bibnamefont {Nakai}}, \bibinfo
  {author} {\bibfnamefont {T.}~\bibnamefont {Shibauchi}}, \bibinfo {author}
  {\bibfnamefont {K.}~\bibnamefont {Behnia}}, \bibinfo {author} {\bibfnamefont
  {K.}~\bibnamefont {Izawa}}, \bibinfo {author} {\bibfnamefont
  {M.}~\bibnamefont {Hedo}}, \bibinfo {author} {\bibfnamefont {Y.}~\bibnamefont
  {Uwatoko}}, \bibinfo {author} {\bibfnamefont {T.}~\bibnamefont {Matsumoto}},
  \bibinfo {author} {\bibfnamefont {R.}~\bibnamefont {Settai}}, \bibinfo
  {author} {\bibfnamefont {Y.}~\bibnamefont {\={O}nuki}}, \bibinfo {author}
  {\bibfnamefont {H.}~\bibnamefont {Kontani}}, \ and\ \bibinfo {author}
  {\bibfnamefont {Y.}~\bibnamefont {Matsuda}},\ }\href@noop {} {\bibfield
  {journal} {\bibinfo  {journal} {J. Phys. Soc. Jpn.}\ }\textbf {\bibinfo
  {volume} {76}},\ \bibinfo {pages} {024703} (\bibinfo {year}
  {2007})}\BibitemShut {NoStop}%
\bibitem [{\citenamefont {Kontani}(2008)}]{KontaniRepProg}%
  \BibitemOpen
  \bibfield  {author} {\bibinfo {author} {\bibfnamefont {H.}~\bibnamefont
  {Kontani}},\ }\href@noop {} {\bibfield  {journal} {\bibinfo  {journal} {Rep.
  Prog. Phys.}\ }\textbf {\bibinfo {volume} {71}},\ \bibinfo {pages} {026501}
  (\bibinfo {year} {2008})}\BibitemShut {NoStop}%
\bibitem [{\citenamefont {Abrikosov}(1998)}]{AbrikosovPRB}%
  \BibitemOpen
  \bibfield  {author} {\bibinfo {author} {\bibfnamefont {A.~A.}\ \bibnamefont
  {Abrikosov}},\ }\href@noop {} {\bibfield  {journal} {\bibinfo  {journal}
  {Phys. Rev. B}\ }\textbf {\bibinfo {volume} {58}},\ \bibinfo {pages} {2788}
  (\bibinfo {year} {1998})}\BibitemShut {NoStop}%
\bibitem [{\citenamefont {Parish}\ and\ \citenamefont
  {Littlewood}(2003)}]{ParishNature}%
  \BibitemOpen
  \bibfield  {author} {\bibinfo {author} {\bibfnamefont {M.~M.}\ \bibnamefont
  {Parish}}\ and\ \bibinfo {author} {\bibfnamefont {P.~B.}\ \bibnamefont
  {Littlewood}},\ }\href@noop {} {\bibfield  {journal} {\bibinfo  {journal}
  {Nature}\ }\textbf {\bibinfo {volume} {426}},\ \bibinfo {pages} {162}
  (\bibinfo {year} {2003})}\BibitemShut {NoStop}%
\bibitem [{\citenamefont {Abrikosov}(2000)}]{AbrikosovEPL}%
  \BibitemOpen
  \bibfield  {author} {\bibinfo {author} {\bibfnamefont {A.~A.}\ \bibnamefont
  {Abrikosov}},\ }\href@noop {} {\bibfield  {journal} {\bibinfo  {journal}
  {Europhys. Lett.}\ }\textbf {\bibinfo {volume} {49}},\ \bibinfo {pages} {789}
  (\bibinfo {year} {2000})}\BibitemShut {NoStop}%
\bibitem [{\citenamefont {Novoselov}\ \emph {et~al.}(2005)\citenamefont
  {Novoselov}, \citenamefont {Geim}, \citenamefont {Morozov}, \citenamefont
  {Jiang}, \citenamefont {Katsnelson}, \citenamefont {Grigorieva},
  \citenamefont {Dubonos},\ and\ \citenamefont {Firsov}}]{NovoselovNature}%
  \BibitemOpen
  \bibfield  {author} {\bibinfo {author} {\bibfnamefont {K.~S.}\ \bibnamefont
  {Novoselov}}, \bibinfo {author} {\bibfnamefont {A.~K.}\ \bibnamefont {Geim}},
  \bibinfo {author} {\bibfnamefont {S.~V.}\ \bibnamefont {Morozov}}, \bibinfo
  {author} {\bibfnamefont {D.}~\bibnamefont {Jiang}}, \bibinfo {author}
  {\bibfnamefont {M.~I.}\ \bibnamefont {Katsnelson}}, \bibinfo {author}
  {\bibfnamefont {I.~V.}\ \bibnamefont {Grigorieva}}, \bibinfo {author}
  {\bibfnamefont {S.~V.}\ \bibnamefont {Dubonos}}, \ and\ \bibinfo {author}
  {\bibfnamefont {A.~A.}\ \bibnamefont {Firsov}},\ }\href@noop {} {\bibfield
  {journal} {\bibinfo  {journal} {Nature}\ }\textbf {\bibinfo {volume} {438}},\
  \bibinfo {pages} {197} (\bibinfo {year} {2005})}\BibitemShut {NoStop}%
\bibitem [{\citenamefont {Taskin}\ \emph {et~al.}(2011)\citenamefont {Taskin},
  \citenamefont {Ren}, \citenamefont {Sasaki}, \citenamefont {Segawa},\ and\
  \citenamefont {Ando}}]{TaskinPRL}%
  \BibitemOpen
  \bibfield  {author} {\bibinfo {author} {\bibfnamefont {A.~A.}\ \bibnamefont
  {Taskin}}, \bibinfo {author} {\bibfnamefont {Z.}~\bibnamefont {Ren}},
  \bibinfo {author} {\bibfnamefont {S.}~\bibnamefont {Sasaki}}, \bibinfo
  {author} {\bibfnamefont {K.}~\bibnamefont {Segawa}}, \ and\ \bibinfo {author}
  {\bibfnamefont {Y.}~\bibnamefont {Ando}},\ }\href@noop {} {\bibfield
  {journal} {\bibinfo  {journal} {Phys. Rev. Lett.}\ }\textbf {\bibinfo
  {volume} {107}},\ \bibinfo {pages} {016801} (\bibinfo {year}
  {2011})}\BibitemShut {NoStop}%
\bibitem [{\citenamefont {Xu}\ \emph {et~al.}(1997)\citenamefont {Xu},
  \citenamefont {Husmann}, \citenamefont {Rosenbaum}, \citenamefont {Saboungi},
  \citenamefont {Enderby},\ and\ \citenamefont {Littlewood}}]{XuNature}%
  \BibitemOpen
  \bibfield  {author} {\bibinfo {author} {\bibfnamefont {R.}~\bibnamefont
  {Xu}}, \bibinfo {author} {\bibfnamefont {A.}~\bibnamefont {Husmann}},
  \bibinfo {author} {\bibfnamefont {T.~F.}\ \bibnamefont {Rosenbaum}}, \bibinfo
  {author} {\bibfnamefont {M.~L.}\ \bibnamefont {Saboungi}}, \bibinfo {author}
  {\bibfnamefont {J.~E.}\ \bibnamefont {Enderby}}, \ and\ \bibinfo {author}
  {\bibfnamefont {P.~B.}\ \bibnamefont {Littlewood}},\ }\href@noop {}
  {\bibfield  {journal} {\bibinfo  {journal} {Nature}\ }\textbf {\bibinfo
  {volume} {390}},\ \bibinfo {pages} {57} (\bibinfo {year} {1997})}\BibitemShut
  {NoStop}%
\bibitem [{\citenamefont {Katayama}\ \emph {et~al.}(2007)\citenamefont
  {Katayama}, \citenamefont {Kobayashi},\ and\ \citenamefont
  {Suzumura}}]{Kobayashijpsj}%
  \BibitemOpen
  \bibfield  {author} {\bibinfo {author} {\bibfnamefont {S.}~\bibnamefont
  {Katayama}}, \bibinfo {author} {\bibfnamefont {A.}~\bibnamefont {Kobayashi}},
  \ and\ \bibinfo {author} {\bibfnamefont {Y.}~\bibnamefont {Suzumura}},\
  }\href@noop {} {\bibfield  {journal} {\bibinfo  {journal} {J. Phys. Soc.
  Jpn.}\ }\textbf {\bibinfo {volume} {75}},\ \bibinfo {pages} {054705}
  (\bibinfo {year} {2007})}\BibitemShut {NoStop}%
\bibitem [{\citenamefont {Park}\ \emph {et~al.}(2011)\citenamefont {Park},
  \citenamefont {Lee}, \citenamefont {Wolff-Fabris}, \citenamefont {Koh},
  \citenamefont {Eom}, \citenamefont {Kim}, \citenamefont {Farhan},
  \citenamefont {Jo}, \citenamefont {Kim}, \citenamefont {Shim},\ and\
  \citenamefont {Kim}}]{ParkPRL}%
  \BibitemOpen
  \bibfield  {author} {\bibinfo {author} {\bibfnamefont {J.}~\bibnamefont
  {Park}}, \bibinfo {author} {\bibfnamefont {G.}~\bibnamefont {Lee}}, \bibinfo
  {author} {\bibfnamefont {F.}~\bibnamefont {Wolff-Fabris}}, \bibinfo {author}
  {\bibfnamefont {Y.~Y.}\ \bibnamefont {Koh}}, \bibinfo {author} {\bibfnamefont
  {M.~J.}\ \bibnamefont {Eom}}, \bibinfo {author} {\bibfnamefont {Y.~K.}\
  \bibnamefont {Kim}}, \bibinfo {author} {\bibfnamefont {M.~A.}\ \bibnamefont
  {Farhan}}, \bibinfo {author} {\bibfnamefont {Y.~J.}\ \bibnamefont {Jo}},
  \bibinfo {author} {\bibfnamefont {C.}~\bibnamefont {Kim}}, \bibinfo {author}
  {\bibfnamefont {J.~H.}\ \bibnamefont {Shim}}, \ and\ \bibinfo {author}
  {\bibfnamefont {J.~S.}\ \bibnamefont {Kim}},\ }\href@noop {} {\bibfield
  {journal} {\bibinfo  {journal} {Phys. Rev. Lett.}\ }\textbf {\bibinfo
  {volume} {107}},\ \bibinfo {pages} {126402} (\bibinfo {year}
  {2011})}\BibitemShut {NoStop}%
\bibitem [{\citenamefont {Wang}\ \emph {et~al.}(2011)\citenamefont {Wang},
  \citenamefont {Graf}, \citenamefont {Lei}, \citenamefont {Tozer},\ and\
  \citenamefont {Petrovic}}]{WangPRB}%
  \BibitemOpen
  \bibfield  {author} {\bibinfo {author} {\bibfnamefont {K.}~\bibnamefont
  {Wang}}, \bibinfo {author} {\bibfnamefont {D.}~\bibnamefont {Graf}}, \bibinfo
  {author} {\bibfnamefont {H.}~\bibnamefont {Lei}}, \bibinfo {author}
  {\bibfnamefont {S.~W.}\ \bibnamefont {Tozer}}, \ and\ \bibinfo {author}
  {\bibfnamefont {C.}~\bibnamefont {Petrovic}},\ }\href@noop {} {\bibfield
  {journal} {\bibinfo  {journal} {Phys. Rev. B}\ }\textbf {\bibinfo {volume}
  {84}},\ \bibinfo {pages} {220401} (\bibinfo {year} {2011})}\BibitemShut
  {NoStop}%
\bibitem [{\citenamefont {Kuo}\ \emph {et~al.}(2011)\citenamefont {Kuo},
  \citenamefont {Chu}, \citenamefont {Riggs}, \citenamefont {Yu}, \citenamefont
  {McMahon}, \citenamefont {De~Greve}, \citenamefont {Yamamoto}, \citenamefont
  {Analytis},\ and\ \citenamefont {Fisher}}]{KuoPRB}%
  \BibitemOpen
  \bibfield  {author} {\bibinfo {author} {\bibfnamefont {H.~H.}\ \bibnamefont
  {Kuo}}, \bibinfo {author} {\bibfnamefont {J.~H.}\ \bibnamefont {Chu}},
  \bibinfo {author} {\bibfnamefont {S.~C.}\ \bibnamefont {Riggs}}, \bibinfo
  {author} {\bibfnamefont {L.}~\bibnamefont {Yu}}, \bibinfo {author}
  {\bibfnamefont {P.~L.}\ \bibnamefont {McMahon}}, \bibinfo {author}
  {\bibfnamefont {K.}~\bibnamefont {De~Greve}}, \bibinfo {author}
  {\bibfnamefont {Y.}~\bibnamefont {Yamamoto}}, \bibinfo {author}
  {\bibfnamefont {J.~G.}\ \bibnamefont {Analytis}}, \ and\ \bibinfo {author}
  {\bibfnamefont {I.~R.}\ \bibnamefont {Fisher}},\ }\href@noop {} {\bibfield
  {journal} {\bibinfo  {journal} {Phys. Rev. B}\ }\textbf {\bibinfo {volume}
  {84}},\ \bibinfo {pages} {054540} (\bibinfo {year} {2011})}\BibitemShut
  {NoStop}%
\bibitem [{\citenamefont {Analytis}\ \emph {et~al.}(2009)\citenamefont
  {Analytis}, \citenamefont {McDonald}, \citenamefont {Chu}, \citenamefont
  {Riggs}, \citenamefont {Bangura}, \citenamefont {Kucharczyk}, \citenamefont
  {Johannes},\ and\ \citenamefont {Fisher}}]{AnalytisPRB}%
  \BibitemOpen
  \bibfield  {author} {\bibinfo {author} {\bibfnamefont {J.~G.}\ \bibnamefont
  {Analytis}}, \bibinfo {author} {\bibfnamefont {R.~D.}\ \bibnamefont
  {McDonald}}, \bibinfo {author} {\bibfnamefont {J.-H.}\ \bibnamefont {Chu}},
  \bibinfo {author} {\bibfnamefont {S.~C.}\ \bibnamefont {Riggs}}, \bibinfo
  {author} {\bibfnamefont {A.~F.}\ \bibnamefont {Bangura}}, \bibinfo {author}
  {\bibfnamefont {C.}~\bibnamefont {Kucharczyk}}, \bibinfo {author}
  {\bibfnamefont {M.}~\bibnamefont {Johannes}}, \ and\ \bibinfo {author}
  {\bibfnamefont {I.~R.}\ \bibnamefont {Fisher}},\ }\href@noop {} {\bibfield
  {journal} {\bibinfo  {journal} {Phys. Rev. B}\ }\textbf {\bibinfo {volume}
  {80}},\ \bibinfo {pages} {064507} (\bibinfo {year} {2009})}\BibitemShut
  {NoStop}%
\bibitem [{\citenamefont {Tsukada}\ \emph {et~al.}(2011)\citenamefont
  {Tsukada}, \citenamefont {Hanawa}, \citenamefont {Komiya}, \citenamefont
  {Ichinose}, \citenamefont {Akiike}, \citenamefont {Imai},\ and\ \citenamefont
  {Maeda}}]{TsukadaJPSJ}%
  \BibitemOpen
  \bibfield  {author} {\bibinfo {author} {\bibfnamefont {I.}~\bibnamefont
  {Tsukada}}, \bibinfo {author} {\bibfnamefont {M.}~\bibnamefont {Hanawa}},
  \bibinfo {author} {\bibfnamefont {S.}~\bibnamefont {Komiya}}, \bibinfo
  {author} {\bibfnamefont {A.}~\bibnamefont {Ichinose}}, \bibinfo {author}
  {\bibfnamefont {T.}~\bibnamefont {Akiike}}, \bibinfo {author} {\bibfnamefont
  {Y.}~\bibnamefont {Imai}}, \ and\ \bibinfo {author} {\bibfnamefont
  {A.}~\bibnamefont {Maeda}},\ }\href@noop {} {\bibfield  {journal} {\bibinfo
  {journal} {J. Phys. Soc. Jpn.}\ }\textbf {\bibinfo {volume} {80}},\ \bibinfo
  {pages} {023712} (\bibinfo {year} {2011})}\BibitemShut {NoStop}%
\bibitem [{\citenamefont {Ran}\ \emph {et~al.}(2009)\citenamefont {Ran},
  \citenamefont {Wang}, \citenamefont {Zhai}, \citenamefont {Vishwanath},\ and\
  \citenamefont {Lee}}]{RanPRB}%
  \BibitemOpen
  \bibfield  {author} {\bibinfo {author} {\bibfnamefont {Y.}~\bibnamefont
  {Ran}}, \bibinfo {author} {\bibfnamefont {F.}~\bibnamefont {Wang}}, \bibinfo
  {author} {\bibfnamefont {H.}~\bibnamefont {Zhai}}, \bibinfo {author}
  {\bibfnamefont {A.}~\bibnamefont {Vishwanath}}, \ and\ \bibinfo {author}
  {\bibfnamefont {D.~H.}\ \bibnamefont {Lee}},\ }\href@noop {} {\bibfield
  {journal} {\bibinfo  {journal} {Phys. Rev. B}\ }\textbf {\bibinfo {volume}
  {79}},\ \bibinfo {pages} {014505} (\bibinfo {year} {2009})}\BibitemShut
  {NoStop}%
\bibitem [{\citenamefont {Tanabe}\ \emph {et~al.}(2011)\citenamefont {Tanabe},
  \citenamefont {Huynh}, \citenamefont {Heguri}, \citenamefont {Mu},
  \citenamefont {Urata}, \citenamefont {Xu}, \citenamefont {Nouchi},
  \citenamefont {Mitoma},\ and\ \citenamefont {Tanigaki}}]{TanabePRBCoexis}%
  \BibitemOpen
  \bibfield  {author} {\bibinfo {author} {\bibfnamefont {Y.}~\bibnamefont
  {Tanabe}}, \bibinfo {author} {\bibfnamefont {K.~K.}\ \bibnamefont {Huynh}},
  \bibinfo {author} {\bibfnamefont {S.}~\bibnamefont {Heguri}}, \bibinfo
  {author} {\bibfnamefont {G.}~\bibnamefont {Mu}}, \bibinfo {author}
  {\bibfnamefont {T.}~\bibnamefont {Urata}}, \bibinfo {author} {\bibfnamefont
  {J.}~\bibnamefont {Xu}}, \bibinfo {author} {\bibfnamefont {R.}~\bibnamefont
  {Nouchi}}, \bibinfo {author} {\bibfnamefont {N.}~\bibnamefont {Mitoma}}, \
  and\ \bibinfo {author} {\bibfnamefont {K.}~\bibnamefont {Tanigaki}},\
  }\href@noop {} {\bibfield  {journal} {\bibinfo  {journal} {Phys. Rev. B}\
  }\textbf {\bibinfo {volume} {84}},\ \bibinfo {pages} {100508} (\bibinfo
  {year} {2011})}\BibitemShut {NoStop}%
\bibitem [{\citenamefont {Tanabe}\ \emph {et~al.}(2012)\citenamefont {Tanabe},
  \citenamefont {Huynh}, \citenamefont {Urata}, \citenamefont {Heguri},
  \citenamefont {Mu}, \citenamefont {Xu}, \citenamefont {Nouchi},\ and\
  \citenamefont {Tanigaki}}]{TanabePRBSuppre}%
  \BibitemOpen
  \bibfield  {author} {\bibinfo {author} {\bibfnamefont {Y.}~\bibnamefont
  {Tanabe}}, \bibinfo {author} {\bibfnamefont {K.~K.}\ \bibnamefont {Huynh}},
  \bibinfo {author} {\bibfnamefont {T.}~\bibnamefont {Urata}}, \bibinfo
  {author} {\bibfnamefont {S.}~\bibnamefont {Heguri}}, \bibinfo {author}
  {\bibfnamefont {G.}~\bibnamefont {Mu}}, \bibinfo {author} {\bibfnamefont
  {J.~T.}\ \bibnamefont {Xu}}, \bibinfo {author} {\bibfnamefont
  {R.}~\bibnamefont {Nouchi}}, \ and\ \bibinfo {author} {\bibfnamefont
  {K.}~\bibnamefont {Tanigaki}},\ }\href@noop {} {\bibfield  {journal}
  {\bibinfo  {journal} {Phys. Rev. B}\ }\textbf {\bibinfo {volume} {86}},\
  \bibinfo {pages} {094510} (\bibinfo {year} {2012})}\BibitemShut {NoStop}%
\bibitem [{\citenamefont {Liu}\ \emph {et~al.}(2010)\citenamefont {Liu},
  \citenamefont {Hu}, \citenamefont {Qian}, \citenamefont {Fobes},
  \citenamefont {Mao}, \citenamefont {Bao}, \citenamefont {Reehuis},
  \citenamefont {Kimber}, \citenamefont {Prokeš}, \citenamefont {Matas},
  \citenamefont {Argyriou}, \citenamefont {Hiess}, \citenamefont {Rotaru},
  \citenamefont {Pham}, \citenamefont {Spinu}, \citenamefont {Qiu},
  \citenamefont {Thampy}, \citenamefont {Savici}, \citenamefont {Rodriguez},\
  and\ \citenamefont {Broholm}}]{LiuNatMat}%
  \BibitemOpen
  \bibfield  {author} {\bibinfo {author} {\bibfnamefont {T.~J.}\ \bibnamefont
  {Liu}}, \bibinfo {author} {\bibfnamefont {J.}~\bibnamefont {Hu}}, \bibinfo
  {author} {\bibfnamefont {B.}~\bibnamefont {Qian}}, \bibinfo {author}
  {\bibfnamefont {D.}~\bibnamefont {Fobes}}, \bibinfo {author} {\bibfnamefont
  {Z.~Q.}\ \bibnamefont {Mao}}, \bibinfo {author} {\bibfnamefont
  {W.}~\bibnamefont {Bao}}, \bibinfo {author} {\bibfnamefont {M.}~\bibnamefont
  {Reehuis}}, \bibinfo {author} {\bibfnamefont {S.~A.~J.}\ \bibnamefont
  {Kimber}}, \bibinfo {author} {\bibfnamefont {K.}~\bibnamefont {Prokeš}},
  \bibinfo {author} {\bibfnamefont {S.}~\bibnamefont {Matas}}, \bibinfo
  {author} {\bibfnamefont {D.~N.}\ \bibnamefont {Argyriou}}, \bibinfo {author}
  {\bibfnamefont {A.}~\bibnamefont {Hiess}}, \bibinfo {author} {\bibfnamefont
  {A.}~\bibnamefont {Rotaru}}, \bibinfo {author} {\bibfnamefont
  {H.}~\bibnamefont {Pham}}, \bibinfo {author} {\bibfnamefont {L.}~\bibnamefont
  {Spinu}}, \bibinfo {author} {\bibfnamefont {Y.}~\bibnamefont {Qiu}}, \bibinfo
  {author} {\bibfnamefont {V.}~\bibnamefont {Thampy}}, \bibinfo {author}
  {\bibfnamefont {A.~T.}\ \bibnamefont {Savici}}, \bibinfo {author}
  {\bibfnamefont {J.~A.}\ \bibnamefont {Rodriguez}}, \ and\ \bibinfo {author}
  {\bibfnamefont {C.}~\bibnamefont {Broholm}},\ }\href@noop {} {\bibfield
  {journal} {\bibinfo  {journal} {Nat. Mater.}\ }\textbf {\bibinfo {volume}
  {9}},\ \bibinfo {pages} {718} (\bibinfo {year} {2010})}\BibitemShut {NoStop}%
\bibitem [{\citenamefont {Kawasaki}\ \emph {et~al.}(2012)\citenamefont
  {Kawasaki}, \citenamefont {Deguchi}, \citenamefont {Demura}, \citenamefont
  {Watanabe}, \citenamefont {Okazaki}, \citenamefont {Ozaki}, \citenamefont
  {Yamaguchi}, \citenamefont {Takeya},\ and\ \citenamefont
  {Takano}}]{KawasakiSSC}%
  \BibitemOpen
  \bibfield  {author} {\bibinfo {author} {\bibfnamefont {Y.}~\bibnamefont
  {Kawasaki}}, \bibinfo {author} {\bibfnamefont {K.}~\bibnamefont {Deguchi}},
  \bibinfo {author} {\bibfnamefont {S.}~\bibnamefont {Demura}}, \bibinfo
  {author} {\bibfnamefont {T.}~\bibnamefont {Watanabe}}, \bibinfo {author}
  {\bibfnamefont {H.}~\bibnamefont {Okazaki}}, \bibinfo {author} {\bibfnamefont
  {T.}~\bibnamefont {Ozaki}}, \bibinfo {author} {\bibfnamefont
  {T.}~\bibnamefont {Yamaguchi}}, \bibinfo {author} {\bibfnamefont
  {H.}~\bibnamefont {Takeya}}, \ and\ \bibinfo {author} {\bibfnamefont
  {Y.}~\bibnamefont {Takano}},\ }\href@noop {} {\bibfield  {journal} {\bibinfo
  {journal} {Solid State Commun.}\ }\textbf {\bibinfo {volume} {152}},\
  \bibinfo {pages} {1135} (\bibinfo {year} {2012})}\BibitemShut {NoStop}%
\bibitem [{\citenamefont {Miyake}\ \emph {et~al.}(2010)\citenamefont {Miyake},
  \citenamefont {Nakamura}, \citenamefont {Arita},\ and\ \citenamefont
  {Imada}}]{MiyakeJPSJ}%
  \BibitemOpen
  \bibfield  {author} {\bibinfo {author} {\bibfnamefont {T.}~\bibnamefont
  {Miyake}}, \bibinfo {author} {\bibfnamefont {K.}~\bibnamefont {Nakamura}},
  \bibinfo {author} {\bibfnamefont {R.}~\bibnamefont {Arita}}, \ and\ \bibinfo
  {author} {\bibfnamefont {M.}~\bibnamefont {Imada}},\ }\href@noop {}
  {\bibfield  {journal} {\bibinfo  {journal} {J. Phys. Soc. Jpn.}\ }\textbf
  {\bibinfo {volume} {79}},\ \bibinfo {pages} {044705} (\bibinfo {year}
  {2010})}\BibitemShut {NoStop}%
\end{thebibliography}%

\end{document}